\documentclass[11pt,onecolumn, draftcls]{IEEEtran}
\usepackage{graphicx}
\usepackage{amsmath}
\usepackage{amssymb}
\usepackage{boxedminipage}

\newtheorem{theorem}{Theorem}
\newtheorem{corollary}{Corollary}
\newtheorem{lemma}{Lemma}
\newtheorem{definition}{Definition}
\newtheorem{property}{Property}

\begin{document}

\title{Optimal Design of Multiple Description Lattice Vector Quantizers}

\author{\authorblockN{Xiang Huang and Xiaolin Wu}\\
\thanks{This work was presented in part at the 2006 Data Compression
Conference, and supported by Natural Sciences and Engineering
Research Council of Canada.}
\thanks{The authors are with the Department of
Electrical \& Computer Engineering, McMaster University, Canada, L8S
4K1 (e-mail: huangx4@mcmaster.ca; xwu@ece.mcmaster.ca)} }

\maketitle

\begin{abstract}

In the design of multiple description lattice vector quantizers
(MDLVQ), index assignment plays a critical role.  In addition, one
also needs to choose the Voronoi cell size of the central lattice
$\nu$, the sublattice index $N$, and the number of side descriptions
$K$ to minimize the expected MDLVQ distortion, given the total
entropy rate of all side descriptions $R_t$ and description loss
probability $p$.  In this paper we propose a linear-time MDLVQ index
assignment algorithm for any $K\geq 2$ balanced descriptions in any
dimensions, based on a new construction of so-called $K$-fraction
lattice.  The algorithm is greedy in nature but is proven to be
asymptotically ($N \to \infty$) optimal for any $K \geq 2$ balanced
descriptions in any dimensions, given $R_t$ and $p$. The result is
stronger when $K=2$: the optimality holds for finite $N$ as well,
under some mild conditions. For $K > 2$, a local adjustment
algorithm is developed to augment the greedy index assignment, and
conjectured to be optimal for finite $N$.

Our algorithmic study also leads to better understanding of $\nu$,
$N$ and $K$ in optimal MDLVQ design. For $K=2$ we derive, for the
first time, a non-asymptotical closed form expression of the
expected distortion of optimal MDLVQ in $p$, $R_t$, $N$. For $K>2$,
we tighten the current asymptotic formula of the expected
distortion, relating the optimal values of $N$ and $K$ to $p$ and
$R_t$ more precisely.

\end{abstract}


\pagebreak

\section{Introduction}
\label{sec:Introduction}

Recent years have seen greatly increased research activities on
multiple description coding (MDC), which are motivated by
cooperative and distributed source coding for network
communications.  In a packet-switched network such as the Internet,
an MDC-coded signal is transmitted in multiple descriptions (called
side descriptions) via different routes from one or multiple servers
to a receiver. Each side description can be independently decoded to
reconstruct the signal at certain fidelity, while multiple side
descriptions can be jointly decoded to reconstruct the signal at
higher fidelity.
By utilizing path diversity (the ability to communicate a content
over different paths from a server to a client) and server diversity
(the possibility of transmitting a source from multiple servers),
MDC codes can weather adverse network conditions much better than
single description codes, particularly in real-time communications
where retransmission is not an option.

Multiple description codes can be generated by three categories of
techniques: quantization, correlating transforms
and erasure correction coding \cite{Goyal2001}.
This paper is concerned with the approach of multiple description
lattice vector quantization \cite{SVS1999, SVS2000, SVS2001,
SVS2002, Goyal2000, Goyal2002, Tian2005, JJR2005, JJR2006}.

The first practical design of multiple description quantizer was the
multiple description scaler quantizer (MDSQ) proposed by
Vaishampayan in $1993$ \cite{Vaishampayan93}. The key mechanism of
Vaishampayan's technique is an index assignment (IA) scheme. In the
case of two descriptions, the IA scheme labels each codeword of
central quantizer by an ordered pair of indices, one for each side
quantizer.  MDSQ first quantizes a signal sample to a central
quantizer codeword, then maps, via index assignment, this codeword
to a pair of side quantizer indices.
Vaishampayan proposed few index assignments for two-description
balanced MDSQ \cite{Vaishampayan93}.  These index assignments,
although asymptotically good, were shown by Berger-Wolf and Reingold
to be suboptimal \cite{IA_KDSQ2002}.  The authors alternatively
formulated MDSQ IA as a combinatorial optimization problem of
arranging consecutive integers in a $K$-dimensional matrix similarly
as in graph bandwidth problem \cite{IA_KDSQ2002}.  With this
formulation they proposed a constructive algorithm for MDSQ index
assignment. The resulting index assignment was shown to minimize the
maximum side distortion given central distortion, but only for a
special case of two balanced description MDSQ when the index
assignment matrix has no null elements (i.e., the number of central
codewords is equal to the square of the number of side codewords,
corresponding to having no redundancy in the system).  Moreover, the
technique of optimizing index assignment by arranging integers in a
matrix cannot be extended to multiple description vector
quantization (MDVQ), because no linear ordering of code vectors in
two or higher dimensions can preserve spatial proximity.


Theoretically, MDVQ can achieve the MDC rate distortion bound as
block length approaches infinity.  Unfortunately, optimal MDVQ
design is computationally intractable (optimal single-description VQ
design is already NP-hard \cite{NPHard}).  A practical way of
managing the complexity is to use lattice VQ codebooks.  This
reduces the MDVQ design problem to one of choosing a lattice
$\Lambda $ for central description and an associated sublattice
$\Lambda _s $ for $K\geq2$ side descriptions, and establishing a
one-to-one mapping, called index assignment $\alpha$, between a
point $\lambda \in \Lambda $ and an ordered $K$-tuple $(\lambda _1
,...,\lambda _K )\in \Lambda_s^K $. The above MDVQ scheme was first
proposed by Servetto {\it et al.} \cite{SVS1999}, and commonly
referred to as multiple description lattice vector quantization
(MDLVQ).  Given the dimension of source vectors, lattices $\Lambda $
and $\Lambda _s$ can be selected from the known optimal and/or
near-optimal lattice vector quantizers (e.g., those tabulated in
\cite{SloaneBook}). Therefore, the key issue in optimal MDLVQ design
is to find the bijection function $\alpha: \Lambda \leftrightarrow
\alpha(\Lambda)\subset \Lambda _s^K $ that minimizes a distortion
measure weighted over all possible channel/network scenarios.


The seminal paper of \cite{SVS2001} studied the index assignment
problem for $K=2$ balanced MDLVQ in considerable length, and
proposed a ``guiding principle'' for constructing an optimal index
assignment for two balanced descriptions. Also, the authors pointed
out that optimal MDLVQ index assignment is a problem of linear
assignment.  However, a challenging algorithmic problem remains.
This is how to reduce the graph matching problem from an association
between two infinite sets $\Lambda$ and $\Lambda_s^K$ to between a
finite subset of $\Lambda$ and a finite subset of $\Lambda_s^K$, and
keep these two finite sets as small as possible without compromising
optimality.

Diggavi {\it et al.} proposed a technique of converting the index
assignment problem for two description lattice VQ to a finite
bipartite graph matching problem \cite{SVS2002}.  Two sublattices
$\Lambda_1$, $\Lambda_2$, and their product sublattice of
$\Lambda_s$ are used to construct the two description LVQ.  The
index assignment is obtained by a minimum weight matching between a
Voronoi set of central lattice points and a set of edges (ordered
pairs of sublattice points, one end point in $\Lambda_1$ and the
other in $\Lambda_2$). Each set has a cardinality of $N_1 N_2$,
where $N_k$ is the index of $\Lambda_k$, $k=1,2$.  Therefore, the
index assignment can be computed in $O((N_1 N_2)^{5/2})$ time, given
that the weighted bipartite graph matching can be solved in
$O(N^{5/2})$ time \cite{Hopcroft}.

In \cite{SVS2002} the authors only argued their index assignment
algorithm to be optimal for two description lattice scalar
quantizers, and left its optimality for lattice vector quantizers
unexamined.  This technique of constructing MDLVQ using a product
sublattice was extended from two descriptions to any $K$ balanced
descriptions by {\O}stergaard {\it et al.} \cite{JJR2006}.
{\O}stergaard {\it et al.} also used linear assignment to find index
assignments.  Their solution seemed to require $O(N^5)$ time, where
$N$ is the sublattice index, because it used a candidate set of
$O(N^2)$ central lattice points.  Even with such a large set of
candidate central lattice points, still no bound was given on the
size of the candidate $K$-tuples of sublattice points used for
labeling, and no proof of optimality was offered.

In this paper we propose an $O(N)$ greedy index assignment algorithm
for MDLVQ of any $K \geq 2$ balanced descriptions in any dimensions.
We prove that the algorithm minimizes the expected distortion given
the loss probability $p$ and entropy rate $R_s$ of side
descriptions, as $N \to \infty$.  Moreover, for $K=2$, we can prove,
under some mild conditions, the optimality of the algorithm for
finite $N$ as well. For $K > 2$ and a finite $N$, we augment the
greedy algorithm by a fast local adjustment procedure, if necessary.
We conjecture that this augmented algorithm is optimal in general.

The remainder of the paper is structured as follows.  The next
section formulates the optimal MDLVQ design problem and introduces
necessary notations. Section \ref{sec:IAAlgorithm} presents the
greedy index assignment algorithm.  An asymptotical ($N \to \infty$)
optimality of the proposed algorithm is proven in Section
\ref{sec:OptimalN}. Constructing the proof leads to some new and
improved closed form expressions of the expected MDLVQ distortion in
$N$ and $K$, which are also presented in the section.  Section
\ref{sec:Non-asymp-Opt} sharpens some results of the previous
section for two balanced descriptions, by proving the optimality and
deriving an exact distortion formula of the proposed algorithm for
finite $N$.  The non-asymptotical results of Section
\ref{sec:Non-asymp-Opt} use a so-called $S$-similar sublattice.
Section \ref{sec:S-similar} shows that common lattices in signal
quantization do have $S$-similar sublattices.  Considering that the
greedy index assignment may be suboptimal for finite $N$ when $K >
2$, we develop in Section \ref{augment} a local adjustment algorithm
to augment it.  Section \ref{sec:Conclusions} concludes.

\section{Preliminaries}
\label{sec:Preliminaries}

In a $K$-description MDLVQ, an input vector $x\in R^L$ is first
quantized to its nearest lattice point $\lambda \in \Lambda$, where
$\Lambda$ is a fine lattice. Then the lattice point $\lambda$ is
mapped by a bijective labeling function $\alpha$ 
to an ordered $K$-tuple $(\lambda_1, \lambda_2, \cdots, \lambda_K)
\in \Lambda_s^K$, where $\Lambda_s$ is a coarse lattice.  Let the
components of $\alpha$ be $(\alpha_1, \alpha_2, \cdots, \alpha_K)$,
i.e., $\alpha_k(\lambda) = \lambda_k$, $1 \leq k \leq K$.  With the
function $\alpha$ the encoder generates $K$ descriptions of $x$:
$\lambda_k$, $1 \leq k \leq K$, and transmits each description via
an independent channel to a receiver.

If the decoder receives all $K$ descriptions, it can reconstruct $x$
to $\lambda$ with the inverse labeling function $\alpha^{-1}$.
In general, due to channel losses, the decoder receives only a
subset $\chi$ of the $K$ descriptions, then it can reconstruct $x$
to the average of the received descriptions:
\begin{equation} \nonumber
\hat x = \frac{1}{|\chi|} \sum_{\lambda_i \in \chi} \lambda_i.
\end{equation}
Note the optimal decoder that minimizes the mean square error should
decode $x$ to the centroid of the points $\lambda \in \Lambda$ whose
corresponding components $\alpha(\lambda)$ are in $\chi$. But
decoding to the average of received descriptions is easy for design
\cite{JJR2005}. It is also asymptotically optimal for two
description case \cite{SVS2001}.



\subsection{Lattice and Sublattice}
A lattice $\Lambda$ in the $L$-dimensional Euclidean space is a
discrete set of points
\begin{equation}
\Lambda  \triangleq \{ \lambda  \in \mathbb{R}^L :\lambda  = uG, u
\in \mathbb{Z}^L \},
\end{equation}
i.e., the set of all possible integral linear combinations of the
rows of a matrix $G$. The $L \times L$ matrix $G$ of full rank is
called a generator matrix for the lattice. The Voronoi cell of a
lattice point $\lambda \in \Lambda$ is defined as
\begin{equation}
V(\lambda ) \triangleq \{ x \in \mathbb{R}^L : \| {x - \lambda } \|
\leqslant \|x - \tilde \lambda \|,\forall \tilde \lambda  \in
\Lambda \},
\end{equation}
where $\left\| x \right\|^2 = \langle x, x \rangle$ is the
dimension-normalized norm of vector $x$.

\begin{figure}[htb]
\centering
  \includegraphics[width=2.5in]{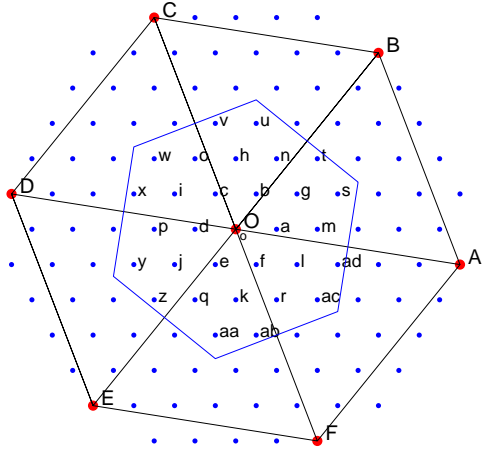}\\
  \caption[Hexagonal lattice $A_2$ and its sublattice with index $N=31$.]{Hexagonal lattice $A_2$ and its sublattice with index $N=31$. Central lattice points are marked by small
dots, and sublattice points by big dots.}
  \label{fig:N31}
\end{figure}

Two lattices are used in the MDLVQ system: a fine lattice $\Lambda$
and a coarse lattice $\Lambda_s$.  The fine lattice $\Lambda$ is the
codebook for the central decoder when all the descriptions are
received, thus called central lattice.  The coarse lattice
$\Lambda_s$ is the codebook for a side decoder when only one
description is received.  Typically, $\Lambda_s \subset \Lambda$,
hence $\Lambda_s$ is also called a sublattice.  The ratio of the
point densities of $\Lambda$ and $\Lambda_s$, which is also the
ratio of the volumes of the Voronoi cells of $\Lambda_s$ and
$\Lambda$, is defined as the sublattice index $N$.  If the
sublattice is clean (no central lattice points lie on the boundary
of a sublattice Voronoi cell), $N$ is equal to the number of central
lattice points inside a sublattice Voronoi cell. Sublattice index
$N$ governs trade-offs between the side and central distortions. We
assume that $\Lambda_s$ is geometrically similar to $\Lambda$, i.e.,
$\Lambda_s$ can be obtained by scaling, rotating, and possibly
reflecting $\Lambda$ \cite{SloaneBook}.
Fig.~\ref{fig:N31} is an example of hexagonal lattice and its
sublattice with index $N=31$.

Let $G$ and $G_s$ be generator matrices for $L$-dimensional central
lattice $\Lambda$ and sublattice $\Lambda_s$. Then $\Lambda_s$ is
geometrically similar to $\Lambda$ if and only if there exist an
invertible $L \times L$ matrix $U$ with integer entries, a scalar
$\beta$, and an orthogonal $L \times L$ matrix $A$ with determinant
$1$ such that
\begin{equation}
\label{eq:similarity}
 G_s = UG = \beta G A,
\end{equation}
The index for a geometrically similar lattice is $N = \frac {det
G_s} {det G} = \beta^L$.

\subsection{Rate of MDLVQ}

In MDLVQ, a source vector $x$ of joint pdf $g(x)$ is quantized to
its nearest fine lattice $\lambda \in \Lambda$. The probability of
quantizing $x$ to a lattice $\lambda$ is
\begin{equation}
P(\lambda) = {\int_{V(\lambda )}} g(x)dx.
\end{equation}
The entropy rate per dimension of the output of the central
quantizer is \cite{SVS2001}
\begin{equation}
\begin{split}
R_c &= \frac{1} {L} \sum_{\lambda \in \Lambda} P(\lambda) \log P(\lambda)\\
&= - \frac{1} {L}\sum\limits_{\lambda  \in \Lambda }
{\int_{V(\lambda )} {g(x)dx} \log _2 \int_{V(\lambda )}
{g(x)dx}}\\
&\approx  - \frac{1} {L}\sum\limits_{\lambda  \in \Lambda }
{\int_{V(\lambda )} {g(x)dx} \log _2 g(\lambda)\nu} \\
&= h(p) - \frac{1} {L} \log _2 \nu. \label {eq:R},
\end{split}
\end{equation}
where $\nu$ is the volume of a Voronoi cell of $\Lambda$, and $h(p)$
is the differential entropy.  The above assumes high resolution when
$g(x)$ is approximately constant within a Voronoi cell $V(\lambda)$.

The volume of a Voronoi cell of the sublattice $\Lambda_s$ is
${\nu}_s = N\nu$.  Denote by $Q(x) = \lambda$ the quantization
mapping.  Then, similarly to (\ref{eq:R}), the entropy rate per
dimension of a side description (for balanced MDLVQ) is
\cite{SVS2001}
\begin{equation}
\begin{split}
R_s &= \frac{1} {L} H(\alpha_k (Q(X))) \\
 &\approx h(p) - \frac{1} {L} \log _2 {\nu}_s\\
 &=h(p) - \frac{1} {L} \log _2 (N\nu).
 \label{eq:Rs}
\end{split}
\end{equation}

The total entropy rate per dimension for the balanced MDLVQ system
is
\begin{equation}
R_t = K R_s.
\end{equation}

\subsection{Distortion of MDLVQ}
Assuming that the $K$ channels are independent and each has a
failure probability $p$, we can write the expected distortion as
\begin{equation} \nonumber
\label{eq:Doriginal} D = \sum_{k = 0}^{K} {K \choose k} (1 - p)^k
p^{K - k} D_k,
\end{equation}
where $D_k$ is the expected distortion when receiving $k$ out of $K$
descriptions.

For the case of all descriptions received, the average distortion
per dimension is given by
\begin{equation}
d_c =\sum\limits_{\lambda \in \Lambda } {\int_{V(\lambda )} {\left\|
{x-\lambda } \right\|^2g(x)dx} } \approx G_{\Lambda} \nu
^{\frac{2}{L}}, \label{eq:dc}
\end{equation}
where $G_{\Lambda}$ is the dimensionless normalized second moment of
lattice $\Lambda$ \cite{SloaneBook}.  The approximation is under the
standard high resolution assumption.

If only description $i$ is received, the expected side distortion is
\cite{SVS2001}
\begin{equation}
\begin{split}
d_i &= \sum\limits_{\lambda \in \Lambda } {\int_{V(\lambda )}
{\left\| {x-\lambda_i} \right\|^2g(x)dx} }\\
&= \sum\limits_{\lambda \in \Lambda } {\int_{V(\lambda )} \left(
\left\| {x-\lambda} \right\|^2 + \left\| {\lambda-\lambda_i} \right\|^2 + 2\langle x-\lambda, \lambda-\lambda_i \rangle \right)g(x)dx}  \\
&\approx d_c + \sum\limits_{\lambda \in \Lambda } {\left\| {\lambda
-\lambda_i} \right\|^2P(\lambda )}, \ \ 1 \leq i \leq K
\end{split}
\end{equation}
Hence the expected distortion when receiving only one description is
\begin{equation}
\begin{split}
D_1 &= \frac{1} {K} \sum _{i=1}^{K}d_i = d_c + \sum\limits_{\lambda
\in \Lambda }\frac{1} {K} \sum\limits_{i=1}^{K}
 {\left\| {\lambda -\lambda_i} \right\|^2P(\lambda )}.
 \label{eq:D1org}
\end{split}
\end{equation}
Let $m_K$ be the centroid of all $K$ descriptions $\lambda_1,
\lambda_2, \cdots, \lambda_K$, that it,
\begin{equation}
m_K \triangleq \frac{1}{K} \sum_{k=1}^{K} \lambda_k.
\end{equation}
Then we have
\begin{equation}
\begin{split}
\frac{1} {K}\sum\limits_{i = 1}^K {\left\| {\lambda  - \lambda _i }
\right\|^2 } &= \frac{1} {K}\sum\limits_{i = 1}^K {\left\| {(\lambda
- m_K  ) - (\lambda _i  - m_K  )} \right\|^2 }\\
&= \left\| {\lambda  - m_K  } \right\|^2  + \frac{1}
{K}\sum\limits_{i = 1}^K {\left\| {\lambda _i  - m_K  } \right\|^2 }
- \frac{2} {K}\left\langle {\lambda  - m_K ,\sum\limits_{i = 1}^K
{(\lambda _i  - m_K  )} } \right\rangle
\\
&= \left\| {\lambda  - m_K  } \right\|^2  + \frac{1}
{K}\sum\limits_{i = 1}^K {\left\| {\lambda _i  - m_K  } \right\|^2
}.
\end{split}
\label{eq:parell}
\end{equation}
Substituting (\ref{eq:parell}) into (\ref{eq:D1org}), we get
\begin{equation}
D_1 = d_c + \sum\limits_{\lambda \in \Lambda} \left( \left\|
{\lambda - m_K  } \right\|^2  +  \frac{1} {K} \sum\limits_{i=1}^{K}
 \left\| {\lambda_i -m_K} \right\|^2 \right) P(\lambda).
 \label{eq:D1}
\end{equation}

Now we consider the case of receiving $k$ descriptions, $1<k<K$.
Let $\rm I$ be the set of all possible combinations of receiving $k$
out of $K$ descriptions.
Let $\iota = (\iota_1, \iota_2, \cdots, \iota_k)$ be an element of
$\rm I$. Under high resolution assumption, we have \cite{JJR2006}
\begin{equation}
\begin{split}
D_k &= d_c + \left| \rm I \right|^{-1} \sum_{\lambda \in \Lambda}
\sum_{\iota \in {\rm I}} \left\| \lambda - \frac{1}{k}
{\sum_{j=1}^{k} \lambda_{\iota_j}} \right\|^2 P(\lambda)\\
&= d_c +  \sum\limits_{\lambda \in \Lambda} \left( \left\| {\lambda
- m_K } \right\|^2  + \frac{K - k} {(K - 1)k} \frac{1}
{K}\sum\limits_{i = 1}^K {\left\| {\lambda _i - m_K } \right\|^2 }
\right) P(\lambda), \ \ 1<k<K. \label{eq:Dk}
\end{split}
\end{equation}

Substituting the expressions of $D_k$ into (\ref{eq:Doriginal}), we
arrive at
\begin{equation}
D = (1 - p^K )d_c  + \sum_{\lambda  \in \Lambda } {\left( {\zeta_1
\left\| {\lambda  - m_K } \right\|^2  + \zeta _2 \frac{1} {K}\sum_{i
= 1}^K {\left\| {\lambda_i  - m_K } \right\|^2 } } \right)}
P(\lambda) + p^K E[\left\| X \right\|^2 ], \label{eq:D}
\end{equation}
where
\begin{eqnarray}
\begin{split}
\zeta_1 &= \sum_{k=1}^{K-1} {K \choose k} (1 - p)^k p^{K-k} = 1- p^K - (1-p)^K\\
\zeta_2 &= \sum_{k=1}^{K - 1} {K \choose k} (1 - p )^k  p^{K-k}
\frac{K-k} {(K - 1)k}. \label{z1z2}
\end{split}
\end{eqnarray}


\subsection{Optimal MDLVQ Design}
Given source and channel statistics and given total entropy rate
$R_t$, optimal MDLVQ design involves (i) the choice of the central
lattice $\Lambda$ and the sublattice $\Lambda_s$; (ii) the
determination of optimal number of descriptions $K$ and of the
optimal sublattice index value $N$; and (iii) the optimization of
index assignment function $\alpha$ once (i) and (ii) are fixed.  We
defer the discussions of optimal values of $K$ and $N$ to Section
\ref{sec:OptimalN}, and first focus on the construction of optimal
index assignment.  It turns out that our new constructive approach
will lead to improved analytical results of $K$ and $N$ in optimal
MDLVQ design.

With fixed $p$, $K$, $\Lambda$, $\Lambda_s$, the optimal MDLVQ
design problem (i.e., minimizing (\ref{eq:D})) reduces to finding
the optimal index assignment $\alpha$ that minimizes the average
side distortion
\begin{equation}
\label{eq:ds} d_s  \triangleq  \sum_{\lambda  \in \Lambda } {\left(
\frac{1} {K}\sum_{i = 1}^K {\left\| \alpha_i(\lambda) -
\mu(\alpha(\lambda)) \right\|^2 } + \zeta \left\| \lambda  -
\mu(\alpha(\lambda)) \right\|^2 \right)} P(\lambda ),
\end{equation}
where
\begin{eqnarray}
\mu(\alpha(\lambda)) = K^{-1} \sum_{i=1}^K \alpha_i(\lambda) \nonumber \\
\begin{split}
\zeta = \frac{\zeta_1}{\zeta_2} = \frac{\sum_{k=1}^{K-1} {K \choose
k} (1 - p)^k p^{K-k}} { \sum_{k=1}^{K - 1} {K \choose k} (1 - p )^k
p^{K-k} \frac{K-k} {(K - 1)k}}.
\end{split}
\label{eq:zeta}
\end{eqnarray}

When $K=2$, the objective function can be simplified to
\begin{equation}
\begin{split}
\label{eq:dsKis2} d_s 
&= \sum_{\lambda  \in \Lambda } {\left( \frac{1} {4} {\left\|
{\lambda _1  - \lambda _2} \right\|^2 } + { \left\| {\lambda  -
\frac{\lambda_1+\lambda_2}{2} } \right\|^2 } \right)} P(\lambda ) .
\end{split}
\end{equation}

\section{Index Assignment Algorithm}
\label{sec:IAAlgorithm}

This section presents a new greedy index assignment algorithm for
MDLVQ of $K \geq 2$ balanced descriptions and examines its
optimality. The algorithm is very simple and it henges on an
interesting new notion of $K$-fraction sublattice.  We first define
this $K$-fraction sublattice and reveal its useful properties for
optimizing index assignment.  Then we describe the greedy index
assignment algorithm.

\subsection{$K$-fraction Sublattice}

In the following study of optimal index assignment for $K$ balanced
descriptions, the sublattice
\begin{equation}
\Lambda _{s/K}  \triangleq \frac{1}{K}\Lambda _s = \{ \tau  \in
\mathbb{R}^L :\tau  = \frac{u}{K}G_s, u \in \mathbb{Z}^L \}
\end{equation}
plays an important role, and it will be referred as the $K$-fraction
sublattice hereafter.


The $K$-fraction sublattice $\Lambda _{s/K}$ has the following
interesting relations to $\Lambda$ and $\Lambda_s$.

\begin{property}
\label{prop:mean} $\mu(\alpha(\lambda)) = K^{-1} {\sum_{k=1}^{K}
\alpha_k(\lambda)} $ is an onto (but not one-to-one) map: $\Lambda
_s^K \to \Lambda _{s/K} $.
\end{property}
\begin{proof}
1) $(\lambda _1 ,\lambda _2, \cdots, \lambda_K) \in \Lambda _s^K
\Rightarrow {\sum_{k=1}^{K} \lambda_k} \in \Lambda_s \to K^{-1}
{\sum_{k=1}^{K} \lambda_k} \in \Lambda _{s/K} $; 2) $\forall \tau
\in \Lambda _{s/K}$, let $\lambda _1 = K\tau, \lambda _2 = \cdots =
\lambda_K = 0$, then $\lambda _1 ,\lambda _2, \cdots, \lambda_K \in
\Lambda _s $ and $\mu(\alpha(\lambda))=\tau$.
\end{proof}

This means that the centroid of any $K$-tuples in $\Lambda _s^K $
must be in $\Lambda _{s/K} $, and further $\Lambda _{s/K} $ consists
only of these centroids.

If two $K$-fraction sublattice points $\tau_1, \tau_2$ satisfy
$\tau_1 - \tau_2 \in \Lambda_s$, then we say that $\tau_1$ and
$\tau_2$ are in the same coset with respect to $\Lambda_s$.  Any
$K$-fraction sublattice point belong to one of the cosets.
\begin{property}
$\Lambda_{s/K}$ has, in the $L$-dimensional space, $K^L$ cosets with
respect to $\Lambda_s$. \label{prop:cosets}
\end{property}
\begin{proof} Let $\tau_1, \tau_2$ be two $K$-fraction sublattice
points. $\tau_1, \tau_2$ can be expressed by
\begin{eqnarray}
\nonumber \tau_1 = \frac{u}{K}G_s, \, \, \tau_2 = \frac{v}{K}G_s,
\end{eqnarray}
where $u = (u_1, u_2, \cdots, u_L) \in \mathbb{Z}^L, v = (v_1, v_2,
\cdots, v_L) \in \mathbb{Z}^L$.  Two points $\tau_1$ and $\tau_2$
fall in the same coset with respect to $\Lambda_s$ if and only if
$u_i \equiv v_i \bmod K$ for all $i = 1, 2, \cdots, L$.  The claim
follows since the reminder of division by $K$ takes on $K$ different
values.
\end{proof}

The $K$-fraction sublattice $\Lambda _{s/K} $ partitions the space
into Voronoi cells. Denote the Voronoi cell of a point $\tau \in
\Lambda _{s/K} $ by
\begin{equation} \nonumber
\label{eq:halfsubcell} V_{s/K} (\tau) = \{x:\left\| {x-\tau }
\right\|\le \left\| {x-\tilde {\tau }} \right\|,\forall \tilde {\tau
}\in \Lambda _{s/K} \}.
\end{equation}

\begin{property}
\label{prop:clean} $\Lambda _{s/K} $ is clean, if $\Lambda _s $ is
clean.
\end{property}
\begin{proof}
Assume for a contradiction that there was a point $\lambda \in
\Lambda $ on the boundary of $V_{s/K} (\tau )$ for a $\tau \in
\Lambda _{s/K} $. Scaling both $\lambda $ and $V_{s/K} (\tau )$ by
$K$ places $K\lambda $ on the boundary of $KV_{s/K} (\tau
)=\{Kx:\left\| {Kx-K\tau } \right\|\le \left\| {Kx-K\tilde {\tau }}
\right\|,\forall \tilde {\tau }\in \Lambda _{s/K} \}$. But $K\lambda
$ is a point of $\Lambda $, and $KV_{s/K} (\tau )$ is nothing but
the Voronoi cell $V_s $ of the sublattice point $K\tau \in \Lambda
_s $, or the point $K\lambda \in \Lambda $ lies on the boundary of
$V_s (K\tau )$, contradicting that $\Lambda _s $ is clean.
\end{proof}

\begin{property}
\label{prop:symmetry} Both lattices $\Lambda_s$ and $\Lambda$ are
symmetric about any point $\tau \in \Lambda_{s/2}$.
\end{property}
\begin{proof}
$\forall \tau \in \Lambda_{s/2}$, we have $2\tau \in \Lambda_s$, so
$2\tau - \lambda_s \in \Lambda_s$ holds for $\forall \lambda_s \in
\Lambda_s$; similarly, $\forall \tau \in \Lambda_{s/2}$, we have
$2\tau \in \Lambda$, so $2\tau - \lambda \in \Lambda$ holds for
$\forall \lambda \in \Lambda$.
\end{proof}

\subsection{Greedy Index Assignment Algorithm}
\label{secsub:greedyalgorithm}

Our motive of constructing the $K$-fraction lattice $\Lambda_{s/K}$
is to relate $\Lambda_{s/K}$ to the central lattice $\Lambda$ in
such a way that the two terms of $d_s$ in (\ref{eq:ds}) can be
minimized independently.  This is brought into light by examining
the partition of the space by Voronoi cells of $K$-fraction
sublattice points.  For simplicity, we assume the sublattice
$\Lambda_s$ is clean (if not, the algorithm still works by employing
a rule to break a tie on the boundary of a sublattice Voronoi cell).
According to Property \ref{prop:clean}, no point $\lambda \in
\Lambda $ is on the boundary of any Voronoi cell of $\Lambda_{s/K}$.
Let
\begin{equation}
\ss(\tau) = \{ (\lambda_1, \lambda_2, \cdots, \lambda_K) \in
\Lambda_s^K | \Sigma_{1 \leq k \leq K} \lambda_k/K = \tau \}
\end{equation}
be the set of all ordered $K$-tuples of sublattice points of
centroid $\tau$, and $\tau \in \Lambda_{s/K}$ by Property 1.

In constructing an index assignment, we sort the members of
$\ss(\tau)$ by $\sum_{k=1}^K \| \lambda_k - \tau \|^2$.  From
$\ss(\tau)$ we select the ordered $K$-tuples in increasing values of
$\sum_{k=1}^K (\lambda_k - \tau)^2$ to label the central lattice
points inside the $K$-fraction Voronoi cell $V_{s/K} (\tau )$, until
all $N_\tau = |\Lambda \cap V_{s/K} (\tau )|$ of those central
lattice points are labeled.  It follows from (\ref{eq:ds}) that any
bijective mapping between the $n(\tau)$ central lattice points and
the $N_\tau$ ordered $K$-tuples of sublattice points yields the same
value of $d_s$. Such an index assignment clearly minimizes the
second term of (\ref{eq:ds}), which is the sum of the squared
distances of all central lattice points in Voronoi cell $V_{s/K}
(\tau )$ to the centroid $\tau = \mu(\alpha(\cdot))$. As $N \to
\infty$, the proposed index assignment algorithm also minimizes the
first term of (\ref{eq:ds}) independently. This will be proven with
some additional efforts in Section \ref{sec:OptimalN}.

For the two description case, these $N_\tau$ ordered pairs are
formed by the $N_\tau$ nearest sublattice points to $\tau$ in
$\Lambda_s$ by Property \ref{prop:symmetry}. Note when $\tau \in
\Lambda_s$, the ordered pair $(\tau, \tau)$ should be used to label
$\tau$ itself.

According to Property \ref{prop:cosets}, $\Lambda_{s/K}$ has $K^L$
cosets with respect to $\Lambda_s$ in the $L$-dimensional space, so
there are $K^L$ classes of $V_{s/K} (\tau )$.  We only need to label
one representative out of each class, and cover the whole space by
shifting. Thus it suffices to label a total of $N$ central lattice
points.

\begin{figure}[htb]
\centering
  \includegraphics[width=5in]{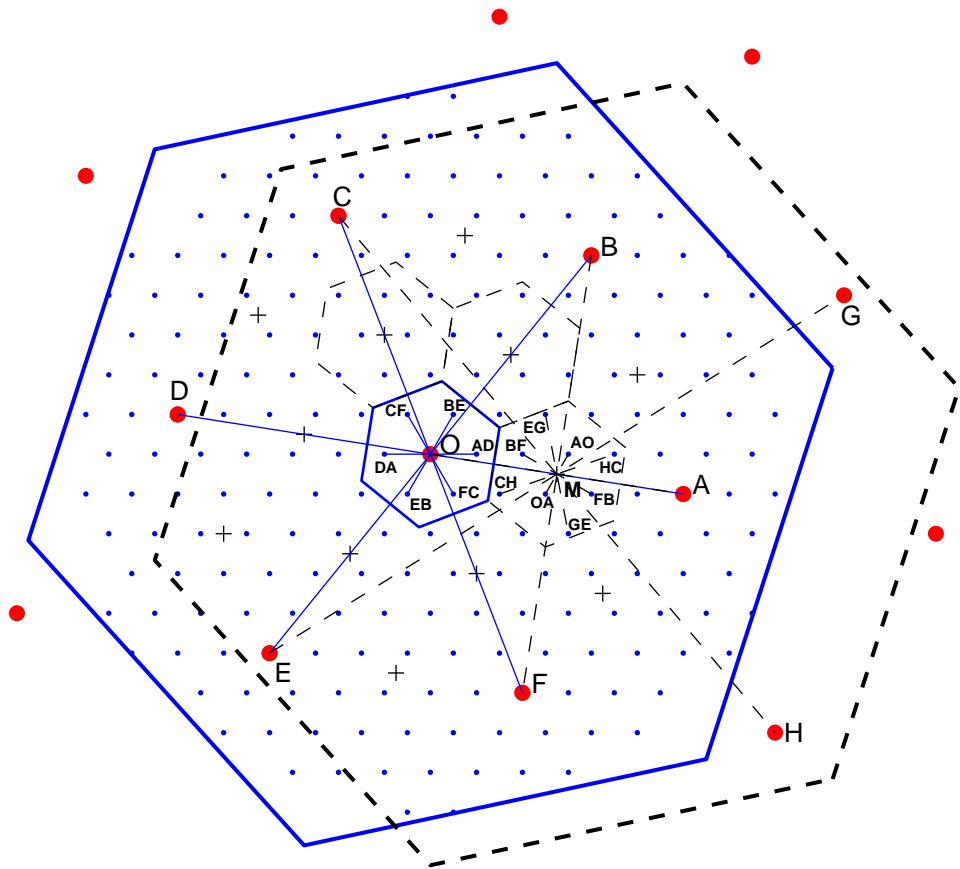}\\
  \caption[Optimal index assignments for $A_2$ lattice with $N=31$, $K=2$]
  {Optimal index assignments for $A_2$ lattice with $N=31$, $K=2$. Points of $\Lambda$, $\Lambda_s$ and $\Lambda_{s/2}$
  are marked by $\cdot$, $\bullet$ and $+$, respectively.}
  \label{fig:N31IA}
\end{figure}

\begin{figure}[htb]
\centering
  \includegraphics[width=5in]{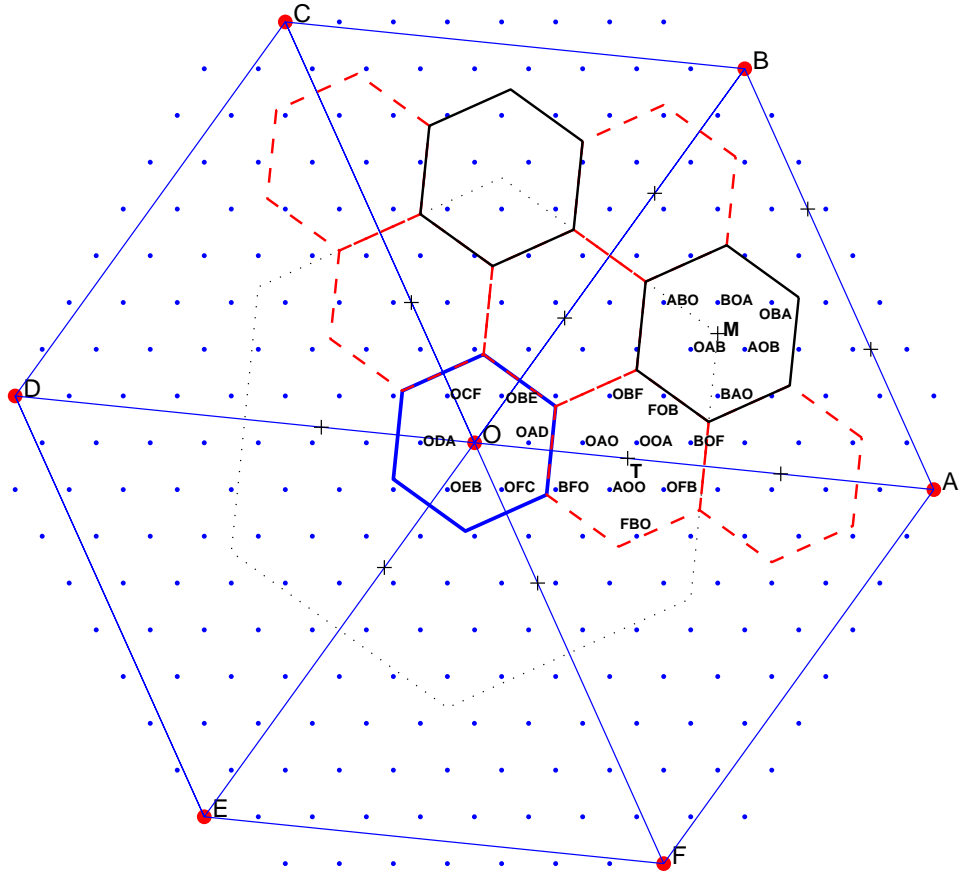}\\
  \caption[Optimal index assignments for $A_2$ lattice with $N=73$, $K=3$.]
  {Optimal index assignments for $A_2$ lattice with $N=73$, $K=3$. Points of $\Lambda$, $\Lambda_s$ and $\Lambda_{s/3}$
  are marked by $\cdot$, $\bullet$ and $+$, respectively.\\}
  \label{fig:N73IA}
\end{figure}


To visualize the work of the proposed index assignment algorithm,
let us examine two examples on an $A_2$ lattice (see
Figs.~\ref{fig:N31IA} and \ref{fig:N73IA}). The $A_2$ lattice
$\Lambda$ is generated by basis vectors represented by complex
numbers: $1$ and $\omega = 1/2 + i{\sqrt 3}/2$. By shifting
invariance of $A_2$ lattice, we only need to label the $N$ central
lattice points that belong to $K^2$ Voronoi cells of
$\Lambda_{s/K}$. By angular symmetry of $A_2$ lattice, we can
further reduce the number of points to be labeled.

The first example is a two-description case, with the sublattice
$\Lambda_s$ given by basis vectors $5-\omega$, $\omega(5-\omega)$,
which is geometrically similar to $\Lambda$, has index $N=31$ and is
clean (refer to Fig.~\ref{fig:N31IA}).
There are two types of Voronoi cells of $\Lambda_{s/2}$, as shown by
the solid and dashed boundaries in Fig.~\ref{fig:N31IA}. The solid
cell is centered at a central lattice point and contains $7$ central
lattice points. The dashed cell is centered at the midpoint of the
line segment $OA$, and contains $8$ central lattice points. To label
the $7$ central lattice points in $V_{s/2}(O)$, we use the $7$
nearest sublattice points to $O$: $(O, A, B, C, D, E, F)$.  They
form $6$ ordered pairs with the midpoint $O$ : $((A, D), (D, A), (B,
E), (E, B), (C, F), (F, C)$, and an unordered pair $(O,O)$ since $O$
is itself a sublattice point.  To label the $8$ central lattice
points in $V_{s/2}(M)$, we use the $8$ nearest sublattice points to
$M$: $(O, A, B, F, C, H, E, G)$.  They form $8$ ordered pairs with
midpoint $M$: $(O,A)$, $(A, O)$, $(B, F)$, $(F, B)$, $(C, H)$, $(H,
C)$, $(E, G)$, $(G, E)$.  The labeling of the $7$ central lattice
points in $V_{s/2}(O)$ and the labeling of the $8$ central lattice
points in $V_{s/2}(M)$ are illustrated in Fig.~\ref{fig:N31IA}.

Fig.~\ref{fig:N73IA} illustrates the result of the proposed
algorithm in the case of three descriptions.  The depicted index
assignment for three balanced descriptions is computed for the
sublattice of index $N=73$ and basis vectors: $8-\omega$,
$\omega(8-\omega)$.

The presented MDLVQ index assignment algorithm is fast with an
$O(N)$ time complexity. The simplicity and low complexity of the
algorithm are due to the greedy optimization approach adopted.



The tantalizing question is, of course, can the greedy algorithm be
optimal?   A quick test on the above two examples may be helpful.
Let the distance between a nearest pair of central lattice points in
$\Lambda$ be one.  For the first example the result of
\cite{SVS2001} (the best so far) is $d_s = 561/31 = 18.0968$, while
the greedy algorithm does better, producing $d _s = 528/31 =
17.0323$. Indeed, in both examples, one can verify that the expected
distortion is minimized as the two terms of $d_s$ in (\ref{eq:ds})
are minimized independently.

\section{Asymptotically Optimal Design of MDLVQ}
\label{sec:OptimalN}

In this section we first prove that the greedy index assignment is
optimal for any $K$, $p$, $\Lambda$ and $\Lambda_s$ as $N
\rightarrow \infty$.  In constructing the proof we derive a close
form asymptotical expression of the expected distortion of optimal
MDLVQ for general $K \geq 2$.  It allows us to determine the optimal
volume of a central lattice Voronoi cell $\nu$, the optimal
sublattice index $N$, and the optimal number of descriptions $K$,
given the total entropy rate of all side descriptions $R_t$ and the
loss probability $p$.  These results, in addition to optimal index
assignment $\alpha$, complete the design of optimal MDLVQ, and they
present an improvement over previous work of \cite{JJR2006}.


\subsection{Asymptotical Optimality of the Proposed Index
Assignment}

Since the second term of $d_s$ is minimized by the Voronoi partition
defined by the $K$-fraction lattice, the optimality of the proposed
index assignment based on the $K$-fraction lattice follows if it
also minimizes the first term of $d_s$.  This is indeed the case
when $N \to \infty$.  To compute the first term of $d_s$, let
\begin{equation} \nonumber
\varsigma_k \triangleq \sum_{i = 1}^{k} \lambda_i, k=1,2,\cdots, K,
\end{equation}
Then
\begin{equation}
\begin{split}
\sum_{k = 1}^K {\left\| {\lambda _k - \tau }\right\|^2 }
&= \sum_{k = 1}^K \left\| \lambda _k - \frac{1}{K}\varsigma_K \right\|^2 \\
&= \left(\sum_{k = 1}^{K-1} \left\| \lambda _k - \frac{1}{K}\varsigma_K \right\|^2 \right)+ \left\| \lambda_K - \frac{1}{K}\varsigma_K \right\|^2\\
&=\left(\sum_{k = 1}^{K-1} \left\| \left(\lambda _k - \frac{1}{K-1}\varsigma_{K-1} \right) + \frac{1}{K-1}\left(\varsigma_{K-1} - \frac{K-1}{K}\varsigma_K \right)\right\|^2 \right)+ \left\| \varsigma_{K-1} - \frac{K-1}{K}\varsigma_K \right\|^2 \\
 &\overset{(a)}{=} \left(\sum_{k = 1}^{K-1} {\left\| {\lambda _k - \frac{1}{K-1}\varsigma_{K-1}} \right\|^2} \right)
 + \frac{K}{K-1} {\left\| {\varsigma_{K-1} - \frac{K-1}{K}\varsigma_K } \right\|^2 } \\
 &\overset{(b)}{=} \sum_{k=1}^{K-1} \frac{k+1}{k} {\left\| \varsigma_k -
\frac{k}{k+1}\varsigma_{k+1} \right\|^2 }.
\end{split}
\label{eq:cost}
\end{equation}
Equality $(a)$ holds because the inner product $\left<\sum_{k =
1}^{K-1} \left( {\lambda _k - \frac{1}{K-1}\varsigma_{K-1}}\right),
\frac{1}{K-1}\left(\varsigma_{K-1} -
\frac{K-1}{K}\varsigma_K\right)\right>$ is zero. After using the
same deduction $K-1$ times, we arrive at equality $(b)$.

Note the one-to-one correspondence between $(\lambda_1, \lambda_2,
\cdots, \lambda_K)$ and $(\varsigma_1, \varsigma_2, \cdots,
\varsigma_K)$.  Also recall that the proposed index assignment uses
the $N_\tau$ (the number of central lattice points in $K$-fraction
Voronoi cell $V_{s/K}(\tau)$) smallest $K$-tuples in $\ss(\tau)$
according to the value of $\sum_{k=1}^K \| \lambda_k - \tau \|^2$.
Finding the $N_\tau$ smallest values of $\sum_{k = 1}^K \| \lambda_k
- \tau \|^2$ in $\ss(\tau)$ is equivalent to finding the $N_\tau$
smallest values of $\sum_{k=1}^{K-1} \frac{k+1}{k} {\left\|
\varsigma_k - \frac{k}{k+1}\varsigma_{k+1} \right\|^2 }$ among the
$(K-1)$-tuples $(\varsigma_1, \varsigma_2, \cdots, \varsigma_{K-1})$
with $\varsigma_K = \sum_{k=1}^K \lambda_k$.

\begin{theorem}
\label{th:optim} The proposed greedy index assignment algorithm is
optimal as $N \to \infty$ for any given $\Lambda$, $\Lambda_s$, $K$,
and $p$.
\end{theorem}
\begin{proof}
The $i^{th}$ nearest sublattice point to
$\frac{k}{k+1}\varsigma_{k+1}$ is approximately on the boundary of
an $L$-dimensional sphere with volume $i N \nu$. Given
$\varsigma_{k+1}$, the $i^{th}$ smallest value of $\| \varsigma_k -
\frac{k}{k+1}\varsigma_{k+1} \|^2$ is approximately $(iN\nu
/B_L)^{\frac{2}{L}} / L = G_L (1+\frac{2}{L}) (iN\nu)^{\frac{2}{L}}
$, where $B_L = G_L^{-\frac{L}{2}}(L+2)^{-\frac{L}{2}}$ is the
volume of an $L$-dimensional sphere of unit radius \cite{SVS2001},
and $G_L$ is the dimensionless normalized second moment of an
$L$-dimensional sphere.

Let $f^{(n)}(\tau)$ be the $n^{th}$ smallest value of $\sum_{k =
1}^K {\left\| {\lambda _k - \tau }\right\|^2 }$ in $\ss(\tau)$ that
is realized at $(\varsigma_1^{(n)}, \varsigma_2^{(n)}, \cdots,
\varsigma_{K-1}^{(n)})$.  Then
\begin{equation}
\begin{split}
f^{(n)} (\tau) &= \sum_{k=1}^{K-1} \frac{k+1}{k} {\left\|
\varsigma_k^{(n)}
- \frac{k}{k+1}\varsigma_{k+1}^{(n)} \right\|^2 }\\
&\approx G_L \left(1 + \frac{2}{L}\right) (N\nu)^{\frac{2}{L}}
\sum_{k=1}^{K-1} \frac{k+1}{k} (i_k^{(n)})^{\frac{2}{L}},
\label{eq:fn}
\end{split}
\end{equation}
in which $(i_1^{(n)}, i_2^{(n)}, \cdots, i_{K-1}^{(n)}) \in
\mathbb{Z}^{K-1}$ is where the sum $\sum_{k=1}^{K-1} \frac{k+1}{k}
(i_k)^{\frac{2}{L}}$ takes on its $n^{th}$ smallest value over all
$(K-1)$-tuples of positive integers.

When $N \to \infty$, the proposed index assignment algorithm takes
the $N_\tau \approx N / K^L$ smallest terms of $\sum_{k = 1}^K
{\left\| {\lambda _k - \tau }\right\|^2 }$ in $\ss(\tau)$ for every
$\tau$.  But (\ref{eq:fn}) states that the $n^{th}$ smallest value
of $\sum_{k = 1}^K {\left\| {\lambda _k - \tau }\right\|^2 }$ is
independent of $\tau$.  Therefore, the first term of $d_s$ is
minimized, establishing the optimality of the resulting index
assignment.
\end{proof}

Remark IV.1: The $O(N)$ MDLVQ index assignment algorithm based on
the $K$-fraction lattice is so far the only one proven to be
asymptotically optimal, except for the prohibitively expensive
linear assignment algorithm.  In the next section, we will
strengthen the above proof in a constructive perspective, and
establish the optimality of the algorithm for finite $N$ when $K=2$.

\subsection{Optimal Design Parameters $\nu$, $N$ and $K$}

Now our attention turns to the determination of the optimal $\nu$
(the volume of a Voronoi cell of $\Lambda$), $N$ (the sublattice
index) and $K$ (the number of descriptions) that achieve minimum
expected distortion, given the total entropy rate of all side
descriptions $R_t$ and loss probability $p$.

Using (\ref{eq:fn}), we have
\begin{equation}
\begin{split}
\label{eq:Kds1} \sum_{\lambda  \in \Lambda } \sum_{k = 1}^K \left\|
\lambda_k - \mu(\alpha(\lambda)) \right\|^2 P(\lambda) &= \sum_{\tau
\in \Lambda_{s/K} }  \sum_{\lambda \in V_{s/K}(\tau)} \sum_{k = 1}^K
\left\| {\lambda _k - \mu(\alpha(\lambda)) } \right\|^2 P(\lambda) \\
&\approx \frac{1}{N_\tau} \sum_{n=1}^{N_\tau}\sum_{k=1}^{K-1} \frac{k+1}{k} {\left\| \varsigma_k^{(n)} - \frac{k}{k+1}\varsigma_{k+1}^{(n)} \right\|^2 } \\
&\approx G_L (1+\frac{2}{L}) \left(N\nu\right)^{\frac{2}{L}}
\frac{1}{N_\tau} \sum_{n=1}^{N_\tau} \sum_{k=1}^{K-1} \frac{k+1}{k}
(i_k^{(n)})^{\frac{2}{L}}
\end{split}
\end{equation}

Consider the region defined as
\begin{equation}
\Omega \triangleq
\left\{ \sum_{k=1}^{K-1} \frac{k+1}{k} x_k^{\frac{2}{L}} \le C
\left| x_1, x_2, \cdots, x_{K-1} \geq 0, \\
x_1, x_2 \cdots, x_{K-1} \in \mathbb{R} \right. \right\}.
\end{equation}
Choose $C$ appropriately so that the volume of $\Omega$ is
$V(\Omega) = N_\tau$. As $N_\tau \rightarrow \infty$, $\Omega$
contains approximately $N_\tau$ optimal integer vectors $(i_1, i_2,
\cdots, i_{K-1})$. These $N_\tau$ points are uniformly distributed
in $\Omega$, with density one point per unit volume. Because the
ratio between the volume occupied by each point and the total volume
is $1/N_{\tau}$, which approaches zero when $N_\tau \rightarrow
\infty$, we can replace the summation by integral and get
\begin{equation}
\begin{split}
\label{eq:Kds2} \frac{1}{N_\tau}\sum_{n=1}^{N_\tau} \sum_{k=1}^{K-1}
\frac{k+1}{k} (i_k^{(n)})^{\frac{2}{L}} &\approx \frac{\int_{x \in
\Omega } \sum_{k=1}^{K-1} \frac{k+1}{k} x_k^{\frac{2}{L}} \,dx }
{\int_{x \in \Omega
} \,dx} \\
&= \frac{\int_{y \in \Omega_0 } \sum_{k=1}^{K-1} y_k^{\frac{2}{L}}
\,dy } {\int_{y \in \Omega_0 } \,dy},
\end{split}
\end{equation}
where $y_k = (\frac{k+1}{k})^{\frac{L}{2}} x_k, k=1,2,\cdots,K-1$,
and $\Omega_0$ is defined as
\begin{equation}
\Omega_0 \triangleq \left\{ \sum_{k=1}^{K-1}  y_k^{\frac{2}{L}} \le C
\left| y_1, y_2, \cdots, y_{K-1} \geq 0, \\
y_1, y_2 \cdots, y_{K-1} \in \mathbb{R} \right. \right\}.
\end{equation}

Substituting (\ref{eq:Kds2}) into (\ref{eq:Kds1}), we have
\begin{equation}
\label{eq:Kds3} \sum_{\lambda  \in \Lambda } \sum_{k = 1}^K \left\|
{\lambda _k - \mu(\alpha(\lambda)) } \right\|^2 P(\lambda) \approx
G_L (1+\frac{2}{L}) \left(N\nu\right)^{\frac{2}{L}} \frac{\int_{y
\in \Omega_0 } \sum_{k=1}^{K-1} y_k^{\frac{2}{L}} \,dy } {\int_{y
\in \Omega_0 } \,dy}.
\end{equation}

Let $V(\Omega_0)$ be the volume of region $\Omega_0$, i.e.,
\begin{equation}
\begin{split}
V(\Omega_0) &= \int_{y \in \Omega_0 } \,dy_1 \,dy_2 \cdots \,dy_{K-1} \\
&= K^{\frac{L}{2}} \int_{x \in \Omega } \,dx_1 \,dx_2 \cdots \,dx_{K-1} \\
&= K^{\frac{L}{2}} N_\tau \\
&= K^{-\frac{L}{2}} N,
\end{split}
\end{equation}
and define the dimensionless normalized $\frac{2}{L}$th moment of
$\Omega_0$:
\begin{equation}
G_{\Omega_0} \triangleq \frac{1}{K-1} \frac{\int_{y \in \Omega_0 }
\sum_{k=1}^{K-1} y_k^{\frac{2}{L}} \,dy } {V(\Omega_0)^{1 +
\frac{2}{L(K-1)}}}.
\end{equation}
Note that scaling $\Omega_0$ does not change $G_{\Omega_0}$.
For the special case $L=1$, the region $\Omega_0$ is a
$(K-1)$-dimensional sphere in the first octant, so the normalized
second moment $G_{\Omega_0} = 4 G_{K-1}$. For the special case
$K=2$, $G_{\Omega_0} = L/(L+2)$ is the normalized $\frac{2}{L}$th
moment of a line $[0, C]$. Generally, using \emph{Dirichilet's
Integral} \cite{Whtl}, we get
\begin{equation}
G_{\Omega_0} = \frac{1}{n + \frac{2}{L}} \frac{\Gamma(\frac{nL}{2}
+1)^{\frac{2}{nL}}} {\Gamma(\frac{L}{2} + 1)^{\frac{2}{L}}}.
\label{eq:G_Omega}
\end{equation}
Hence,
\begin{equation}
\label{eq:Kds4} \frac{\int_{y \in \Omega_0 } \sum_{k=1}^{K-1}
y_k^{\frac{2}{L}} \,dy } {\int_{y \in \Omega_0 } \,dy} =
G_{\Omega_0}(K-1) V(\Omega_0)^{\frac{2}{L(K-1)}} =
G_{\Omega_0}(K-1)K^{\frac{-1}{K-1}} N^{\frac{2}{L(K-1)}}.
\end{equation}

Substituting (\ref{eq:Kds4}) into (\ref{eq:Kds3}), we have
\begin{equation}
\begin{split}
\label{eq:ds_1st} \frac{1}{K}\sum_{\lambda  \in \Lambda } \sum_{k =
1}^K \left\| {\lambda _k - \mu(\alpha(\lambda)) } \right\|^2
P(\lambda)  &\approx G_L G_{\Omega_0} (1 +
\frac{2}{L})(K-1)K^{\frac{-K}{K-1}}
N^{\frac{2K}{L(K-1)}} \nu^{\frac{2}{L}}\\
 &\approx G_L
\Phi_{K-1, L} (K-1)K^{\frac{-K}{K-1}} N^{\frac{2K}{L(K-1)}}
\nu^{\frac{2}{L}},
\end{split}
\end{equation}
where
\begin{equation} \nonumber
\label{eq:Phi} \Phi_{n, L} = \frac{1 + \frac{2}{L}}{n + \frac{2}{L}}
\frac{\Gamma(\frac{nL}{2} +1)^{\frac{2}{nL}}} {\Gamma(\frac{L}{2} +
1)^{\frac{2}{L}}}.
\end{equation}
Note $\Phi_{n, 1} = 12 G_n$ and $\Phi_{1, L} = 1$.

When $N \to \infty$, $N_\tau \approx N/K^L$ independently of the
cell center $\tau$. The $N_\tau$ central lattice points are
uniformly distributed in $V_{s/K}(\tau)$ whose volume is
approximately $N_\tau \nu$. Hence the second term of $d_s$ can be
evaluated as
\begin{equation}
\begin{split}
\label{eq:ds_2nd} \zeta \sum_{\lambda  \in \Lambda } \left\|
{\lambda - \mu(\alpha(\lambda)) } \right\|^2 P(\lambda ) &\approx
\zeta G_{\Lambda} (N_{\tau}\nu)^{\frac{2}{L}}\\ &= \zeta G_{\Lambda}
K^{-2} (N\nu)^{\frac{2}{L}}.
\end{split}
\end{equation}

Comparing (\ref{eq:ds_2nd}) with (\ref{eq:ds_1st}), the first term
of $d_s$ dominates the second term when $N \rightarrow \infty$, thus
\begin{equation}
\label{eq:dsasym} d_s \approx  G_L \Phi_{K-1, L}
(K-1)K^{\frac{-K}{K-1}} N^{\frac{2K}{L(K-1)}} \nu^{\frac{2}{L}}.
\end{equation}

Substituting (\ref{eq:dsasym}) and (\ref{eq:dc}) into (\ref{eq:D}),
we finally express the expected distortion of optimal MDLVQ in a
closed form:
\begin{equation}
D \approx (1 - p^K )G_{\Lambda} {\nu}^{\frac{2}{L}} + \zeta_2 G_L
\Phi_{K-1, L} (K-1)K^{\frac{-K}{K-1}} N^{\frac{2K}{L(K-1)}}
\nu^{\frac{2}{L}} + p^K E[\| X \|^2]. \label{eq:Dasym}
\end{equation}

Using a different index assignment algorithm {\O}stergaard {\it et
al.} derived a similar expression for the expected MDLVQ distortion
(equation (35) in \cite{JJR2006}):
\begin{equation}
D^* \approx (1 - p^K )G_{\Lambda} {\nu}^{\frac{2}{L}} + \hat K G_L
\psi_L^2 N^{\frac{2K}{L(K-1)}} \nu^{\frac{2}{L}} + p^K E[\| X \|^2],
\label{eq:Dasym_JJR}
\end{equation}
where
\begin{equation}
\hat K = \sum_{k=1}^{K - 1} {K \choose k} (1 - p )^k  p^{K-k}
\frac{K-k} {2kK}
\end{equation}
and $\psi_L$ is a quantity that is given analytically only for $K=2$
and for $K=3$ with odd $L$ and is determined empirically for other
cases.

To compare $D$ and $D^*$, we rewrite (\ref{eq:Dasym}) as
\begin{equation}
D \approx (1 - p^K )G_{\Lambda} {\nu}^{\frac{2}{L}} + \hat K G_L
\hat \psi_L^2 N^{\frac{2K}{L(K-1)}} \nu^{\frac{2}{L}} + p^K E[\| X
\|^2], \label{eq:Dasym2}
\end{equation}
where
\begin{equation}
\hat \psi_L = \sqrt{2 K^{\frac{-1}{K-1}} \Phi_{K-1, L}} .
\end{equation}

The two expressions are the same when $K=2$ for which $\hat \psi_L =
\psi_L = 1$, but they differ for $K>2$.  Table \ref{tab:psi} lists
the values of $\psi_L$ and $\hat \psi_L$ for $K=3$, and it shows
that $\hat \psi_{\infty} = \psi_{\infty} =
(\frac{4}{3})^{\frac{1}{4}}$, and $ \hat \psi_L < \psi_L$ for other
values of $L$.  This implies $D < D^*$, or that our index assignment
makes the asymptotical expression of $D$ tighter.

\begin{table}[htb]
\centering
\begin{tabular}{|c|l|l|}
  \hline
  $L$ & $\psi_L$ & $\hat \psi_L...$ \\
  $1$ & $1.1547...$ & $0.9549...$ \\
  $2$ & $1.1481...$ & $0.9428...$ \\
  $3$ & $1.1346...$ & $0.9394...$ \\
  $5$ & $1.1241...$ & $0.9400...$ \\
  $7$ & $1.1173...$ & $0.9431...$ \\
  $9$ & $1.1125...$ & $0.9466...$ \\
  $11$ & $1.1089...$ & $0.9498...$ \\
  $13$ & $1.1060...$ & $0.9527...$ \\
  $15$ & $1.1036...$ & $0.9552...$ \\
  $17$ & $1.1017...$ & $0.9575...$ \\
  $19$ & $1.1000...$ & $0.9596...$ \\
  $21$ & $1.0986...$ & $0.9614...$ \\
  $51$ & $1.0884...$ & $0.9763...$ \\
  $71$ & $1.0856...$ & $0.9807...$ \\
  $101$ & $1.0832...$ & $0.9848...$ \\
  $\infty$ & $1.0746...$ & $1.0746...$ \\
  \hline
\end{tabular}
\caption{Values of $\psi_L$ and $\hat \psi_L$ in $L$ for $K=3$.
Values of $\psi_L$ are reproduced from Table $1$ in \cite{JJR2006}.}
\label{tab:psi}
\end{table}

Now we proceed to derive the optimal value of $N$, which governs the
optimal trade-off between the central and side distortions for given
$p$ and $K$.  For the total target entropy rate $R_t = KR$, we
rewrite (\ref{eq:R}) to get
\begin{equation}
\label{eq:eta} N\nu = 2^{L(h(p) -R_t/K)}.
\end{equation}
For simplicity, define
\begin{equation}
\eta \triangleq 2^{L(h(p) -R_t/K)},
\end{equation}
and we have
\begin{equation}
\label{eq:Dasymnu} D = (1 - p^K )G_{\Lambda} {\nu}^{\frac{2}{L}} +
\zeta_2 G_L \Phi_{K-1, L} (K-1)K^{\frac{-K}{K-1}}
{\eta}^{\frac{2K}{L(K-1)}} \nu^{\frac{-2}{L(K-1)}} + p^K E[\| X
\|^2].
\end{equation}

Differentiating $D$ with respect to $\nu$ yields the optimal $\nu$
value:
\begin{equation} \nonumber
\nu_{opt} = \eta \left( \frac{\zeta_2}{1-p^K}
\frac{G_L}{G_{\Lambda}} \frac{\Phi_{K-1,L}} {K^{\frac{K}{K-1}}}
\right)^{\frac{L(K-1)}{2K}}.
\end{equation}
Substituting $\nu_{opt}$ to (\ref{eq:eta}), we get optimal $N$:
\begin{equation}
N_{opt}= \left( \frac{1-p^K}{\zeta_2} \frac{G_{\Lambda}}{G_L}
 \frac{K^{\frac{K}{K-1}}}{\Phi_{K-1,L}}
\right)^{\frac{L(K-1)}{2K}}.
\end{equation}
If $K=2$, the expression of $N_{opt}$ can be simplified as
\begin{equation}
N_{opt}= \left( \frac{2(1+p)}{p} \frac{G_{\Lambda}}{G_L}
\right)^{\frac{L}{4}}.
\end{equation}

Remark IV.2: $N_{opt}$ is independent of the total target entropy
rate $R_t$ and source entropy rate $h(p)$. It only depends on the
loss probability $p$ and on the number of descriptions $K$.
Substituting $\nu_{opt}$ into (\ref{eq:Dasymnu}), the average
distortion can be expressed as a function of $K$. Then optimal $K$
can be solved numerically.

Remark IV.3: When $K =2$,  (\ref{eq:dsasym}) can be simplified to
\begin{equation}
d_s \approx \frac{1}{4} G_L (N^2\nu)^{\frac{2}{L}}.
\label{eq:dsasymK2}
\end{equation}
For any $a \in (0,1)$, let $N=2^{L(aR+1)}$, then $\nu =
2^{L(h(p)-(a+1)R-1)}$.  Since $R \rightarrow \infty$ implies $N
\rightarrow \infty$, substituting the expressions of $N$ and $\nu$
into (\ref{eq:dc}) and (\ref{eq:dsasymK2}), we get
\begin{equation} \nonumber
\mathop {\lim }\limits_{R \to \infty } d_c 2^{2R(1 + a)}  = \frac{1}
{4}G_{\Lambda}2^{2h(p)}
\end{equation}
\begin{equation} \nonumber
\mathop {\lim }\limits_{R \to \infty } d_k 2^{2R(1 - a)}  = G_L
2^{2h(p)}, \ \ k = 1,2
\end{equation}
Therefore, the proposed MDLVQ algorithm asymptotically achieves the
second-moment gain of a lattice for the central distortion, and the
second-moment gain of a sphere for the side distortion, which is the
same as the expression in \cite{SVS2001}.  In other words, our
algorithm realizes the MDC performance bound for two balanced
descriptions.


\section{Non-asymptotical Optimality for $K=2$}
\label{sec:Non-asymp-Opt}

In this section we sharpen the results of the previous section, by
proving non-asymptotical (i.e., with respect to a finite $N$)
optimality and deriving an exact distortion formula of our MDLVQ
design algorithm for $K=2$ balanced descriptions, under mild
conditions.  The following analysis is constructive and hence more
useful than an asymptotical counterpart because the value of $N$ is
not very large in practice \cite{Goyal2002}.

\subsection{A Non-asymptotical Proof}

Our non-asymptotical proof is built upon the following definitions
and lemmas.


\begin{definition}
A sublattice $\Lambda_s$ is said to be centric, if the sublattice
Voronoi cell $V_s(\lambda)$ centered at $\lambda \in \Lambda_s$
contains the $N$ nearest central lattice points to $\lambda$.
\end{definition}

Figs.~\ref{fig:N31IA} and \ref{fig:N73IA} show two examples of
centric sublattices.

To prove the optimality of the greedy algorithm, we need some
additional properties.

\begin{lemma}
\label{subOrbits} Assume the sublattice $\Lambda_s$ is centric. If
$\lambda \in V_{s/2} (\tau )$ and $\tilde \lambda \not \in V_{s/2}
(\tilde \tau)$, where $\lambda, \tilde \lambda \in \Lambda$ and
$\tau, \tilde \tau \in \Lambda_{s/2}$, then $\| {\lambda -\tau }
\|\le \| {\tilde \lambda -\tilde \tau } \|$.
\end{lemma}
\begin{proof}
Scaling both $\lambda $ and $V_{s/2} (\tau )$ by $2$ places the
lattice point $2\lambda$ in $V_s (2\tau)$; scaling both $\tilde
\lambda $ and $V_{s/2} (\tilde \tau )$ by $2$ places the lattice
point $2 \tilde \lambda \not \in V_s (2\tilde \tau)$. Since a
sublattice Voronoi cell contains the nearest central lattice points,
$\| {2\lambda -2\tau } \|\le \| {2\tilde \lambda -2\tilde {\tau }}
\|$, and hence $\| {\lambda -\tau } \|\le \| {\tilde \lambda -\tilde
\tau } \|$.
\end{proof}

\begin{definition}
A sublattice $\Lambda_s$ is said to be $S$-similar to $\Lambda$, if
$\Lambda_s$ can be generated by scaling and rotating $\Lambda$
around any point $\tau \in \Lambda _{s/2}$ and $\Lambda_s \subset
\Lambda$.
\end{definition}
Note that the $S$-similarity requires that the center of symmetry be
a point in $\Lambda_{s/2}$.

In what follows we assume that sublattice $\Lambda_s$ is $S$-similar
to $\Lambda$.  Also, we denote by $V_{\tau}$ the region created by
scaling and rotating $V_{s/2} (\tau )$ around $\tau$.

\begin{lemma}
\label{isoOrbits} If $\lambda_s \in V_{\tau}$ and $\tilde \lambda_s
\not \in V_{\tilde \tau}$, where $\lambda_s, \tilde \lambda_s \in
\Lambda_s$ and $\tau, \tilde \tau \in \Lambda_{s/2}$, then $\|
{\lambda_s -\tau } \|\le \| {\tilde \lambda_s -\tilde \tau } \|$.
\end{lemma}
\begin{proof}
This lemma follows from Lemma \ref{subOrbits} and the definition of
$S$-Similar.
\end{proof}
\begin{lemma}
\label{nearest}$\forall \tau \in \Lambda_{s/2}$, the sublattice
points in $V_{\tau}$ form $|\Lambda \cap V_{s/2} (\tau )|$ nearest
ordered $2$-tuples with their midpoints being $\tau$.
\end{lemma}
\begin{proof}
Letting $\tilde \tau = \tau$ in Lemma \ref{isoOrbits}, we see that
$V_{\tau}$ contains the $|\Lambda_s \cap V_{\tau}| =|\Lambda \cap
V_{s/2} (\tau )|$ nearest sublattice points to $\tau$. And these
sublattice points are symmetric about $\tau$ according to Property
\ref{prop:symmetry}. Thus this lemma holds.
\end{proof}

\begin{theorem}
\label{th:optim} The proposed index assignment algorithm is optimal
for $K=2$ and any $N$, if the sublattice is centric and $S$-Similar
to the associated central lattice.
\end{theorem}
\begin{proof}
By Property \ref{prop:mean}, for any $\lambda_1,\lambda_2 \in
\Lambda_s$, $(\lambda_1 + \lambda_2)/2 \in \Lambda_{s/2}$.  Now
referring to (\ref{eq:dsKis2}), the proposed algorithm minimizes the
second term $\sum_{\lambda \in \Lambda } {\| {\lambda - (\lambda_1 +
\lambda_2)/2} \|^2 P(\lambda )}$ of $d_s$, since it labels any
central lattice point $\lambda \in V_{s/2}(\tau)$ by
$(\lambda_1,\lambda_2) \in \Lambda_s^2$, and $(\lambda_1 +
\lambda_2)/2 = \tau$.

The algorithm also independently minimizes the first term
$\sum_{\lambda \in \Lambda } \frac {1}{4} {\| {\lambda _1 -\lambda
_2 } \|^2P(\lambda )} $ of $d_s $. Assume that $\sum_{\lambda \in
\Lambda } {\| {\lambda _1 -\lambda _2 } \|^2P(\lambda )} $ was not
minimized. Then there exists an ordered $2$-tuple $(\tilde \lambda
_1, \tilde \lambda _2) \in \Lambda_s^2$ which is not used in the
index assignment, and $\| {\tilde \lambda _1 - \tilde \lambda _2 }
\| < \| {\lambda _1 - \lambda _2} \|$, where $(\lambda _1 ,\lambda
_2 ) \in \Lambda_s^2$ is used in the index assignment.  Let $\tau =
(\lambda_1 + \lambda_2)/2, \tilde \tau = (\tilde \lambda_1 + \tilde
\lambda_2)/2$.  Since $(\lambda _1 , \lambda _2)$ is used to label a
central lattice point in $V_{s/2}(\tau )$, $\lambda _1 , \lambda _2
\in V_{\tau}$ by Lemma \ref{nearest}. However, $\tilde \lambda _1,
\tilde \lambda _2 \not \in V_{\tilde \tau}$, otherwise $(\tilde
\lambda _1, \tilde \lambda _2)$ would be used in the index
assignment by Lemma \ref{nearest}. So we have $\| {\lambda _1 -\tau}
\| \le \| {\tilde \lambda_1 -\tilde \tau} \|$ by Lemma
\ref{isoOrbits}, hence $\| {\lambda _1 -\lambda_2} \| \le \| {\tilde
\lambda_1 -\tilde \lambda_2} \|$, contradicting $\| {\tilde \lambda
_1 - \tilde \lambda _2 } \| < \| {\lambda _1 - \lambda _2} \|$.
\end{proof}

\begin{figure}
  \centering
  \includegraphics[width=5in]{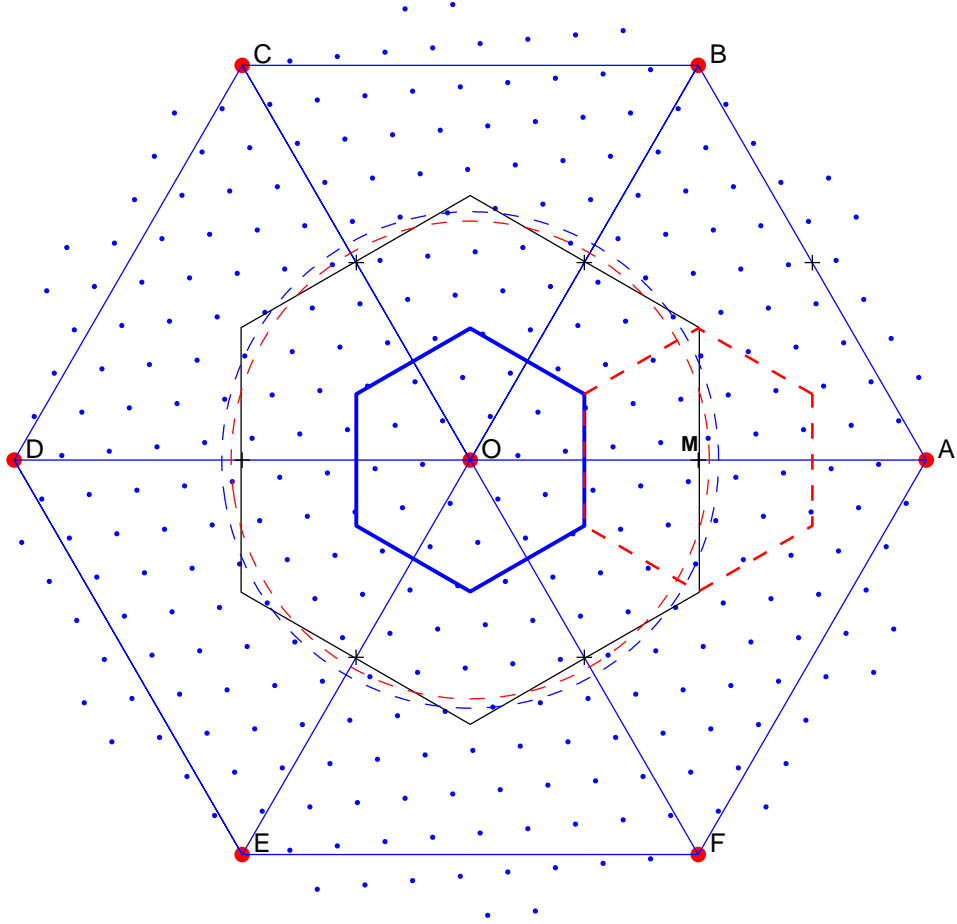}\\
  \caption[The greedy algorithm is optimal for $N=91$,
  although the sublattice is not centric]
  {The greedy algorithm is optimal for $N=91$,
  although the sublattice is not centric.
  The $19$ central lattice points in $V_{s/2}(O)$
  are labeled by the $19$ nearest ordered $2$-tuples with centroid $O$.
  The $24$ central lattice points in $V_{s/2}(M)$
  are labeled by the $24$ nearest ordered $2$-tuples with centroid $M$.
  Let the edge length of $2$-tuple $(O,A)$ be one: $|| O-A|| \triangleq 1$.
  The $19th$ $(20th)$ nearest ordered $2$-tuple with centroid $O$ has
  edge length $4$ $(2\sqrt 7)$. The $24th$ $(25th)$ nearest ordered $2$-tuple
  with centroid $M$ has edge length $5$ $(3 \sqrt 3)$. Because $4 < 3 \sqrt 3$
  and $5<2\sqrt 7$, the first term of $d_s$ is minimized. The second term of
  $d_s$ is minimized because the greedy algorithm partition the space by the
  Voronoi cells of the $K$-fraction sublattice.}
  \label{fig:N91}
\end{figure}

Remark V.1: A sublattice Voronoi cell being centric is not a
necessary condition for the optimality of the greedy algorithm. For
instance, for the $A_2$ lattice generated by basis vectors $1$ and
$\omega = 1/2 + i{\sqrt 3}/2$ and the sublattice of index $N=91$
that is generated by basis vectors $9-\omega$, $\omega(9-\omega)$, a
sublattice Voronoi cell does not contain the $N$ nearest central
lattices, but the greedy algorithm is still optimal as the two terms
of $d_s$ are still independently minimized. This is shown in Figure
\ref{fig:N91}.

Remark V.2: It is easy to choose a centric sublattice for relatively
small $N$ and in high dimensional lattices. For instance, the
sublattices of $A_2$ lattice shown in Fig.~\ref{fig:N31IA},
Fig.~\ref{fig:N73IA} are centric. And any sublattice of $Z$ lattice
is centric.

%
%
%
%

\subsection{Exact Distortion Formula for $K=2$}

We have derived an asymptotical expected distortion formula
(\ref{eq:Dasym}) of the proposed MDLVQ design, which improved a
similar result in \cite{JJR2006}.  But so far no exact
non-asymptotical expression of the expected MDLVQ distortion is
known even for balanced two descriptions.  This subsection presents
a progress on this account.



\begin{lemma}
If the sublattice is clean and $S$-similar, then the second term of
$d_s$ for the proposed optimal MDLVQ design for $K=2$ is
\begin{equation}
\sum_{\lambda  \in \Lambda } {{\left\| {\lambda  - m_{1,2} }
\right\|^2 P(\lambda )} } = \frac{1} {4L} \frac{\sum_{i = 1}^N
a_i}{N}, \label{eq:ds2Analytical}
\end{equation}
where $a_i$ is the squared distance of the $i^{th}$ nearest central
lattice point in $V_s(0)$ to the origin. \label{th:ds2Analytical}
\end{lemma}

\begin{proof}
$\Lambda_{s/2}$ has $2^L$ cosets with respective to $\Lambda_s$ in
the $L$-dimensional space. Let $\tau_1, \tau_2, \cdots, \tau_{2^L}$
be representatives of each coset. For example,  when $L=2$, $\tau_1
= (0, 0)G_s, \tau_2 = (0, \frac{1}{2})G_s, \tau_3 = (\frac{1}{2},
0)G_s, \tau_4 = (\frac{1}{2}, \frac{1}{2})G_s$. Denote by
$V_{\lambda}(\tau) \triangleq V_{s/2}(\tau) \bigcap \Lambda$ the set
of central lattice points in the Voronoi cell of a $2$-fraction
sublattice point $\tau$. We first prove that
\begin{equation}
2\left(V_{\lambda}(\tau_i)-\tau_i\right) \mathop \cap_{i \neq j}
2\left(V_{\lambda}(\tau_j)-\tau_j\right) = \varnothing.
\label{eq:cap}
\end{equation}
\begin{equation}
\mathop \cup _{i=1}^{2^L} 2\left(V_{\lambda}(\tau_i)-\tau_i\right) =
V_s(0)\cap \Lambda. \label{eq:cup}
\end{equation}
Here for convenience, we denote by $2V$ the set of lattice points
that is generated by scaling the lattice points in Voronoi cell $V$
by $2$.


Assume that (\ref{eq:cap}) does not hold. Then there exist
$\lambda_i \in V_{\lambda}(\tau_i), \lambda_j \in
V_{\lambda}(\tau_j)$ such that $\lambda_i - \tau_i = \lambda_j -
\tau_j$. Let $\tau_0 = \tau_i - \tau_j$, then $\tau_0 \in
\Lambda_{s/2}$. We also have $\tau_0 = \lambda_i - \lambda_j$, so
$\tau_0 \in \Lambda$. The sublattice $\Lambda_s$ is $S$-similar to
$\Lambda$, so properly rotating and scaling $\Lambda$ around the
$2$-fraction sublattice point $\tau_0$ can generate $\Lambda_s$.
Rotating and scaling the central lattice point $\tau_0 \in \Lambda$
around $\tau_0$ itself generates $\tau_0$, so $\tau_0 = \tau_j -
\tau_j \in \Lambda_s$.  This contradicts that $\tau_i$ and $\tau_j$
are in different cosets with respect to $\Lambda_s$, establishing
(\ref{eq:cap}).

To prove (\ref{eq:cup}), we first show that for any $\tau \in
\Lambda_{s/2}$,
\begin{equation}
2(V_{\lambda}(\tau)-\tau) = 2\left(V_{s/2}(\tau) \cap \Lambda
\right) -2\tau = V_s(2\tau) \cap (2\Lambda) - 2\tau \overset{(a)}
{\subseteq} V_s(0) \cap \Lambda.
\end{equation}
Step (a) holds because $2\tau  \in \Lambda$ and $V_s(2\tau) - 2\tau
= V_s(0)$. Therefore,
\begin{equation}
\mathop \cup _{i=1}^{2^L} 2\left(V_{\lambda}(\tau_i)-\tau_i\right)
\subseteq V_s(0)\cap \Lambda.
\end{equation}
According to Property \ref{prop:clean}, no central lattice points
lie on the boundary of a $K$-fraction Voronoi cell when the
sublattice is clean, so the set $\mathop \cup _{i=1}^{2^L}
V_{\lambda}(\tau_i)$ contains $N$ different central lattice points.
By (\ref{eq:cap}), the set $\mathop \cup _{i=1}^{2^L}
2\left(V_{\lambda}(\tau_i)-\tau_i\right)$ has $N$ different
elements. Because the set $V_s(0)\cap \Lambda$ also has $N$
different elements and $\mathop \cup _{i=1}^{2^L}
2\left(V_{\lambda}(\tau_i)-\tau_i\right) \subseteq V_s(0)\cap
\Lambda$, (\ref{eq:cup}) holds.

Finally, it follows from (\ref{eq:cap}) and (\ref{eq:cup}) that
\begin{equation}
\begin{split}
\sum_{\lambda  \in \Lambda } {{\left\| {\lambda  -
\mu(\alpha(\lambda)) } \right\|^2 P(\lambda )} } &= \frac{1}{4}\sum
_{\lambda  \in \Lambda } {\left\| {2\lambda -2
\mu(\alpha(\lambda))} \right\|^2 P(\lambda )} \\
&\overset{(a)}{=} \frac{1}{4N}\sum_{i=1}^{4} \sum_{\lambda \in
V_{\lambda}(\tau_i)} {\left\| {2\lambda -2 \tau_i } \right\|^2 }\\
&= \frac{1}{4N} \sum_{\lambda \in V_s(0)\cap \Lambda} {\left\|
\lambda \right\|^2 }\\
&= \frac{1} {4L} \frac{\sum_{i = 1}^N a_i}{N}.
\end{split}
\end{equation}
Equality (a) holds because under high resolution assumption,
$P(\lambda)$ is the same for each central lattice point $\lambda \in
\mathop \cup _{i=1}^{2^L} V_{\lambda}(\tau_i)$.
\end{proof}

\begin{theorem}
If the sublattice is clean, $S$-similar and centric, then the
expected distortion $D$ of optimal two-description MDLVQ is
\begin{equation}
D = (1 - p^2 )G_{\Lambda}\nu ^{\frac{2}{L}}  + \frac{1}{2}p(1-p)
L^{-1}(1+N^{\frac{2}{L}})N^{-1} {\sum_{i = 1}^N a_i} + p^2 E[\left\|
X \right\|^2 ]. \label{eq:Danalytical}
\end{equation}
\end{theorem}
\begin{proof}
By Theorem \ref{th:optim}, under the stated conditions, the proposed
MDLVQ design is optimal.  Further, the corresponding index
assignment makes the first term of $d_s$ exactly $N^{\frac{2}{L}}$
times the second term of $d_s$.  Then it follows from Lemma
\ref{th:ds2Analytical} that the first term of $d_s$ is
\begin{equation}
\sum_{\lambda  \in \Lambda } {\frac{1} {4}\left\| {\lambda _1 -
\lambda _2 } \right\|^2 P(\lambda )}   = \frac{N^{\frac{2}{L}}} {4L}
\frac{\sum_{i = 1}^N a_i} {N}. \label{eq:ds1Analytical}
\end{equation}
Substituting (\ref{eq:dc}), (\ref{eq:ds2Analytical}) and
(\ref{eq:ds1Analytical}) into (\ref{eq:D}), we obtain the formula of
the expected distortion $D$ in (\ref{eq:Danalytical}).
\end{proof}

The above equations lead to some interesting observations.  When the
sublattice is centric, $a_i$ is also the squared distance of the
$i^{th}$ nearest central lattice point to the origin. The term
$N^{\frac{2}{L}} N^{ - 1} \sum_{i = 1}^N a_i$ is the average squared
distance of the $N$ nearest sublattice points to the origin, which
was also realized by previous authors \cite{SVS2001}. The other term
$N^{ - 1} \sum_{i = 1}^N a_i$ is the average squared distance of
central lattice points in $V_s(0)$ to the origin.

The optimal $\nu$ and $N$ for a given entropy rate of side
descriptions can be found by using (\ref{eq:Danalytical}) and $R_s =
h(p) - \frac{1} {L} \log _2 N \nu$ (shown in (\ref{eq:Rs})), rather
than solving many instances of index assignment problem for varying
$N$.


\section{$S$-Similarity}
\label{sec:S-similar}

The above non-asymptotical optimality proof requires the
$S$-similarity of the sublattice. In this section we show that many
commonly used lattices for signal quantization, such as $A_2$, $Z$,
$Z^2$, $Z^L (L=4k )$, and $Z^L$ ($L$ odd), have $S$-similar
sublattices.

Being geometrically similar is a necessary condition of being
$S$-Similar, but being clean is not (For example a geometrically
similar sublattice of $A_2$ with index $21$ is $S$-Similar but not
clean). The geometrical similar and clean sublattices of $A_2$, $Z$,
$Z^2$, $Z^L (L=4k )$, and $Z^L$ ($L$ odd) lattices are discussed in
\cite{SVS2002}. We will discuss the $S$-Similar sublattices of these
lattices in this section.


\begin{theorem}
For the $Z$ lattice $\Lambda$, a sublattice $\Lambda_s$ is
$S$-Similar to $\Lambda$, if and only if its index $N$ is odd.
\end{theorem}
\begin{proof}
Staightforward and omitted.
\end{proof}

\begin{theorem}
\label{th:isoA2} For the $A_2$ lattice $\Lambda$, a sublattice
$\Lambda_s$ is $S$-similar to $\Lambda$, if it is geometrically
similar to $\Lambda$ and clean.
\end{theorem}
\begin{proof}
Let $\Lambda_s$ be a sublattice geometrically similar to $\Lambda$
and clean. We refer to the hexagonal boundary of a Voronoi cell in
$\Lambda$ (respectively in $\Lambda_s$) as $\Lambda$-gon
(respectively $\Lambda_s$-gon).
Any point $\tau \in \Lambda_{s/2}$ is either in $\Lambda_s$ or the
midpoint of a $\Lambda_s$-gon edge.  For instance, in Figure
\ref{fig:N31IA} $M$ is both the midpoint of a $\Lambda$-gon edge and
the midpoint of a $\Lambda_s$-gon edge.

If $\tau \in \Lambda_s$, then $\tau \in \Lambda$, hence scaling and
rotating $\Lambda$ around $\tau$ yields $\Lambda_s$ in this case. If
$\tau$ is the midpoint of a $\Lambda_s$-gon edge, then $\tau \not
\in \Lambda$ because sublattice $\Lambda_s$ is clean, but $\tau \in
\Lambda_{1/2}$, so $\tau$ is the midpoint of a $\Lambda$-gon edge,
hence scaling and rotating $\Lambda$ around $\tau$ yields
$\Lambda_s$ in this case.
\end{proof}


The $Z^L (L=4l, l \geq 1)$ lattice has a geometrically similar and
clean sublattice with index $N$, if and only if $N = m^\frac{L}{2}$,
where $m$ is odd \cite{SVS2002}. Here we show that there are
$S$-similar sublattices for at least half of these $N$ values.
\begin{theorem}
\label{th:isoZL} The $Z^L (L=4l, l \geq 1)$ lattice $\Lambda$ has an
$S$-similar, clean sublattice with index $N$, if $N =
m^{\frac{L}{2}}$ with $m \equiv 1 \bmod 4$.
\end{theorem}

\begin{proof}
We begin with the case $L=4$. By Lagrange's four-square theorem,
there exist four integers $a, b, c, d$ such that $m = a^2 + b^2 +
c^2 + d^2$.  The matrix $G_{\xi}$ constructed by \emph{Lipschitz
integral quaternions} $\{\xi = a + bi + cj +dk\}$ \cite{SVS2002} is


\begin{equation} \nonumber
G_{\xi} = \left(%
\begin{array}{cccc}
   a &  b & c & d \\
  -b &  a & d & -c \\
  -c & -d & a & b \\
  -d & c & -b & a \\
\end{array}%
\right).
\end{equation}
The lattice $\Lambda_s$ generated by matrix $G_s = G_{\xi}$ is a
geometrically similar sublattice of $\Lambda$.

Let $\lambda = u$, $\lambda_s = u_s G_{\xi}$, $\tau =  \frac{1} {2}
u_{\tau} G_{\xi}$ be a point of $\Lambda, \Lambda_s, \Lambda_{s/2}$
respectively, where $u, u_s, u_{\tau} \in \mathbb{Z}^L$. Then,
\begin{equation} \nonumber
\lambda _s  - \tau  = (u_s  - \frac{1} {2}u_\tau  )G_{\xi}.
\end{equation}
Let $\tilde u = u - u_\tau \frac{1} {2}(G_{\xi} - I_L)$, where $I_L$
is an $L \times L$ identity matrix, then
\begin{equation} \nonumber
\lambda - \tau = \tilde u  - \frac{1} {2}u_{\tau}.
\end{equation}
Since $n^2 \equiv 1 \bmod 4$ or $n^2 \equiv 0 \bmod 4$ depending on
whether $n$ is an odd or even integer, $m \equiv 1 \bmod 4$ implies
that exactly one of $a, b, c, d$ is odd. Letting $a$ be odd and
$b,c,d$ even, then $\frac{1} {2}(G_{\xi} - I_L )$ is an integer
matrix. Hence $\tilde u \in \mathbb{Z}^L$. Thus, scaling and
rotating $\Lambda $ around point $\tau$ by scaling factor
$\beta=m^{1/2}$ and rotation matrix $A=m^{-1/2}G_{\xi}$ yields
$\Lambda _s $, proving $\Lambda_s$ is $S$-similar to $\Lambda$.

For the dimension $L = 4l, l>1$, let the $4l \times 4l$ generator
matrix of the sublattice $\Lambda_s$ be
\[
G_s  = \left( {\begin{array}{cccc}
   {G_{\xi}  } & 0 &  \cdots  & 0  \\
   0 & {G_{\xi}  } &  \cdots  &  \vdots   \\
    \vdots  &  \vdots  &  \ddots  & 0  \\
   0 &  \cdots  & 0 & {G_{\xi}  }  \\

 \end{array} } \right).
 \]
Then $\Lambda_s$ is $S$-similar to $\Lambda$. And according to
\cite{SVS2002}, $\Lambda_s$ is clean.
\end{proof}

The $Z_2$ lattice $\Lambda$ has a geometrically similar sublattice
$\Lambda_s$ of index $N$,  if and only if $N = a^2 + b^2, a, b \in
\mathbb{Z}$. And a generator matrix for $\Lambda_s$ is
\begin{equation}
\label{eq:GsforZ2}
 G_s  = \left( {\begin{array}{cccc}
   a & b  \\
   { - b} & a  \\
 \end{array} } \right).
\end{equation}
Further, $\Lambda_s$ is clean if and only if $N$ is odd
\cite{SVS2002}.

\begin{theorem}
For the $Z_2 $ lattice $\Lambda$, a sublattice $\Lambda_s$ is
$S$-similar to $\Lambda$, if it is geometrically similar to
$\Lambda$ and clean.
\end{theorem}
\begin{proof}
For a geometrically similar and clean sublattice $\Lambda_s$, its
generator matrix $G_s$ is given by (\ref{eq:GsforZ2}). As $N=a^2
+b^2$ is odd, $a$ and $b$ are one even and the other odd. Letting
$a$ be odd and $b$ even, by the same argument in proving Theorem
\ref{th:isoZL}, scaling and rotating $\Lambda $ around any point
$\tau \in \Lambda_{s/2}$ by scaling factor $\beta=N^{1/2}$ and
rotation matrix $A=N^{-1/2}G_s$ yields $\Lambda_s$. If $a$ is even,
$b$ is odd, scaling and rotating $\Lambda $ around any point $\tau
\in \Lambda_{s/2}$ by scaling factor $\beta$ and rotation matrix
$\tilde A$ yields $\Lambda _s$, where $\tilde A$ is an orthogonal
matrix:
\[
\tilde A = A\left( {\begin{array}{*{20}c}
   0 & { - 1}  \\
   1 & 0  \\

 \end{array} } \right) = N^{ - 1/2} \left( {\begin{array}{*{20}c}
   b & { - a}  \\
   a & b  \\

 \end{array} } \right).
\]
\end{proof}

\begin{theorem}
\label{th:isoL} An $L$-dimensional lattice $\Lambda$ has an
$S$-similar sublattice with index $N$, if $N = m^L$ is odd.
\end{theorem}
\begin{proof}
Constructing a sublattice $\Lambda_s$ with index $N = m^L$ needs
only scaling, i.e., $G_s = mG$. Let $\lambda = u G$, $\lambda_s = m
u_s G$, $\tau =  \frac{1} {2} m u_{\tau} G$ be in $\Lambda,
\Lambda_s, \Lambda_{s/2}$ respectively, where $u, u_s, u_{\tau} \in
\mathbb{Z}^L$. Then,
\begin{equation} \nonumber
\lambda _s  - \tau  = m(u_s  - \frac{1} {2}u_\tau  )G.
\end{equation}
Let $\tilde u = u - \frac{m-1} {2}u_{\tau}$, then $\tilde u \in
\mathbb{Z}^L$, and
\begin{equation} \nonumber
\lambda - \tau = (\tilde u  - \frac{1} {2}u_{\tau})G.
\end{equation}
Thus, scaling $\Lambda $ around point $\tau$ by  $\beta=m^{1/L}$
yields $\Lambda _s $, proving $\Lambda_s$ is $S$-similar to
$\Lambda$.
\end{proof}

\begin{corollary}
\label{th:isoZL_2} The $Z^L$ ($L$ is odd) lattice $\Lambda$ has an
$S$-similar, clean sublattice with index $N$, if and only if $N =
m^L$ is odd.
\end{corollary}
\begin{proof}
By \cite{SVS2002}, $\Lambda$ has a geometrically similar, clean
sublattice of index $N$, if and only if $N = m^L$ is odd. A
sublattice $\Lambda_s$ of this index can be obtained by scaling
$\Lambda$ by $m$. Theorem \ref{th:isoL} implies that $\Lambda_s$ is
$S$-similar to $\Lambda$.
\end{proof}

\section{Local Adjustment Algorithm}
\label{augment}

\begin{figure}[htb]
\centering
  \includegraphics[width=4in]{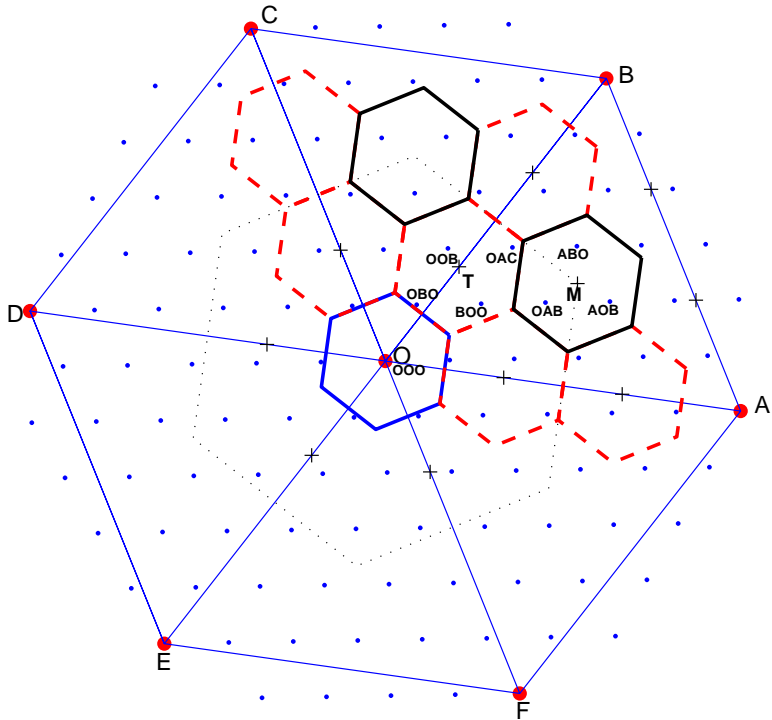}\\
  \caption[Index assignments (not optimal) by the greedy index assignment algorithm for the
  $A_2$ lattice, $N=31$, $K=3$.]{Index assignments (not optimal) by the greedy index assignment algorithm for the
  $A_2$ lattice with index $N=31$, $K=3$. Points of $\Lambda$, $\Lambda_s$ and $\Lambda_{s/3}$
  are marked by $\cdot$, $\bullet$ and $+$, respectively.}
  \label{fig:N31K3IA_bad}
\end{figure}

Theorem \ref{th:optim} is concerned with when the two terms of $d_s$
in (\ref{eq:ds}) can be minimized independently by the greedy index
assignment algorithm.  While being mostly true for $K=2$ as stated
by the theorem and as we saw in Fig. 2 and Fig. 4
\ref{secsub:Examples}, this may not be guaranteed when $K>2$.
Fig.~\ref{fig:N31K3IA_bad} presents the index assignment generated
by the greedy algorithm for $K=3$ on $A_2$ lattice.  The solution is
now suboptimal.  Indeed, consider the central lattice point in
$V_{s/3}(T)$ that is labeled by $OAC$ in Fig.~\ref{fig:N31K3IA_bad},
changing the label from $OAC$ to $BOA$ will reduce $d_s$ of the
central lattice point in question. The change reduces the first term
of $d_s$, although the second term of $d_s$ increases slightly. Note
that the $3$-tuple $(O, A, C)$ has centroid $T$, and the $3$-tuple
$(B, O, A)$ has centroid $M$.

In order to make up for the loss of optimality by the greedy
algorithm, we develop a local adjustment algorithm.  If a central
lattice point $\lambda$ is labeled by an ordered $K$-tuple that has
centroid $\tau \in \Lambda_{s/K}$, we say that $\lambda$ is
attracted by site $\tau$.  If two Voronoi cells $V_{s/K} (\tau_1 )$
and $V_{s/K} (\tau_2 )$ are spatially adjacent, we say that site
$\tau_1$ and site $\tau_2$ are neighbors.  In Figure
\ref{fig:N31K3IA_bad}, site $O$ and site $T$ are neighbors, while
site $O$ and site $M$ are not neighbors.

\begin{figure}
  \centering
  \includegraphics[width=3in]{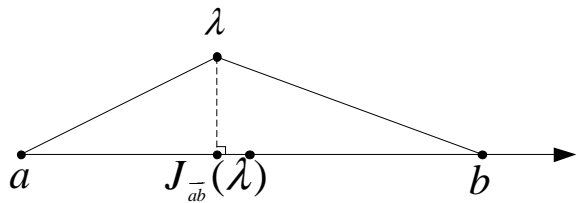}\\
  \caption{Remove lattice $\lambda$ from site $a$, and add it to site $b$}\label{fig:adjust1}
\end{figure}

In Fig.~\ref{fig:adjust1}, assume two neighboring sites $a$ and
$b$ attract $m$ and $n$ central lattice points respectively.  The
$m$ ($n$) central lattice points are labeled by $m$ ($n$) nearest
ordered $K$-tuples centered at site $a$ ($b$).  For any point $x\in
R^L$, let $J_{\overrightarrow{ab}}(x)$ be the projection value of
$x$ onto the axis $\overrightarrow{ab}$.  Consider the set $S(a)$ of
all the $m$ points currently attracted by site $a$, and find
\begin{equation}
\lambda_{max} = \arg \max_{\lambda \in S(a)}
J_{\overrightarrow{ab}}(\lambda).
\end{equation}
Now, introduce an operator $\Rsh(a,b)$ that alters the label of
$\lambda_{max}$ to an ordered $K$-tuple of sublattice points
centered at $b$.  The effect of $\Rsh(a,b)$ is that sites $a$ and
$b$ attract $m-1$ and $n+1$ central lattice points respectively,
which are respectively labeled by $m-1$ and $n+1$ nearest ordered
$K$-tuples centered at site $a$ and site $b$.

>From the definition of side distortion $d_s = \sum_{\lambda \in
\Lambda} d(\lambda) P(\lambda )$ in (\ref{eq:ds}), we have
\begin{equation}
d(\lambda) = \left( \frac{1} {K}\sum_{k = 1}^K {\left\| {\alpha_k
(\lambda) - \mu(\alpha(\lambda)) } \right\|^2 } \right) + \left(
\zeta { \left\| {\lambda - \mu(\alpha(\lambda)) } \right\|^2 }
\right). \label{eq:dlambda}
\end{equation}
Let us compute the change of $d(\lambda_{max})$ caused by the
operation $\Rsh(a,b)$.

The change in the second term of $d(\lambda_{max})$ is
\begin{equation}
\begin{split}
&\zeta \left( { \left\| {\lambda_{max} - b } \right\|^2  - \left\|
{\lambda_{max}  - a
} \right\|^2 } \right)  \\
&=   \frac{\zeta}{L} \left(
\left({J_{\overrightarrow{ab}}(\lambda_{max}) - L \left\| b -a
\right\| }\right)^2 - J_{\overrightarrow{ab}}(\lambda_{max})^2
\right).
\end{split}
\end{equation}
Note the change of the second term is positive if $\lambda_{max} \in
V_{s/K}(a)$.

The change in the first term is
\begin{equation}
f_b(n +1) - f_a(m),
\end{equation}
where $f_\tau(i)$ is the $i^{th}$ smallest value of $\frac{1} {K}
\sum_{k = 1}^K {\left\| {\lambda _k - \tau } \right\|^2 }$ over all
ordered $K$-tuples $(\lambda_1, \lambda_2, \cdots, \lambda_K) \in
\Lambda_s^K$ such that $m(\lambda_1, \lambda_2, \cdots,
\lambda_K)=\tau$.

The net change in $d(\lambda_{max})$ made by operation $\Rsh(a, b)$
is then
\begin{equation}
\Delta (a,b) = \zeta \left( { \left\| {\lambda_{max} - b }
\right\|^2  - \left\| {\lambda_{max}  - a } \right\|^2 } \right) +
f_b(n +1) - f_a(m).
\end{equation}
If $\Delta (a,b) < 0$, then $\Rsh(a,b)$ improves index assignment.

The preceding discussions lead us to a simple local adjustment
algorithm:
\begin{center}
\begin{tabbing}
$(a^*, b^*) = \arg \min_{a \ \mbox{neighbors} \ b} \Delta(a,b)$; \\
While \= $\Delta (a^*,b^*) < 0$ do \\
\> $\Rsh(a,b)$; \\
\> $(a^*, b^*) = \arg \min_{a \ \mbox{neighbors} \ b} \Delta(a,b)$.
\end{tabbing}
\end{center}

Note that it is only necessary to invoke the local adjustment
$\Rsh(a,b)$ if the greedy algorithm does not simultaneously minimize
the two terms of $d_s$.

\begin{figure}[htb]
\centering
  \includegraphics[width=4in]{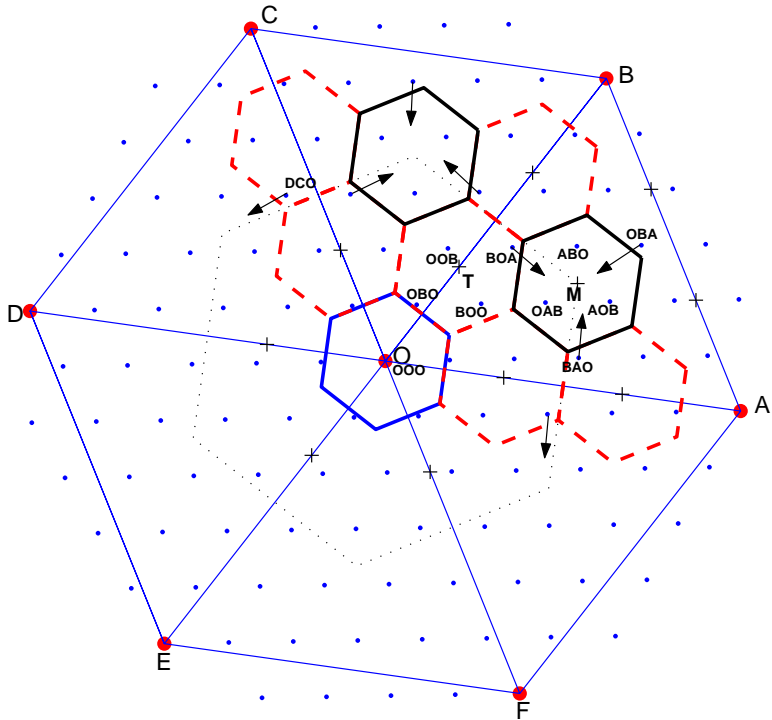}\\
  \caption[Optimal index assignments for the
  $A_2$ lattice, $N=31$, $K=3$.]{Optimal index assignments for the
  $A_2$ lattice, $N=31$, $K=3$. Points of $\Lambda$, $\Lambda_s$ and $\Lambda_{s/3}$
  are marked by $\cdot$, $\bullet$ and $+$, respectively.}
  \label{fig:N31K3IA}
\end{figure}


Fig.~\ref{fig:N31K3IA} shows the result of applying the local
adjustment algorithm to the output of the greedy algorithm presented
in Fig.~\ref{fig:N31K3IA_bad}.  It is easy to prove that the local
adjustment algorithm indeed finds the optimal index assignment for
this case of three description MDLVQ.

Finally, we conjecture that a combined use of the greedy algorithm
and local adjustment $\Rsh(a,b)$ solves the problem of optimal MDLVQ
index assignment for any $L$-dimensional lattice and for all values
of $K$ and $N$.

\section{Conclusion}
\label{sec:Conclusions}

Although optimal MDLVQ index assignment is conceptually a problem of
linear assignment, it involves a bijective mapping between two
infinite sets $\Lambda$ and $\Lambda_s^K$.  No good solutions are
known to reduce the underlying bipartite graph to a modest size
while ensuring optimality.  We developed a linear-time algorithm for
MDLVQ index assignment, and proved it to be asymptotically (in the
sublattice index value $N$) optimal for any $K \geq 2$ balanced
descriptions in any dimensions.  For two balanced descriptions the
optimality holds for finite values of $N$ as well, under some mild
conditions.  We conjecture that the algorithm, with an appropriate
local adjustment, is also optimal for any values of $K$ and $N$.

The optimal index assignment is constructed using a new notion of
$K$-fraction lattice.  The $K$-fraction lattice also lends us a
better tool to analyze and quantify the MDLVQ performance.  The
expected distortion of optimal MDLVQ is derived in exact closed form
for $K=2$ and any $N$.  For cases $K > 2$, we improved the current
asymptotic expression of the expected distortion.  These results can
be used to determine the optimal values of $K$ and $N$ that minimize
the expected MDLVQ distortion, given the total entropy rate and
given the loss probability.

\section*{acknowledgment}
The authors wish to thank Dr. Sorina Dumitrescu for many stimulating
discussions.
%

\bibliographystyle{IEEEtran}
\bibliography{ref}

\end{document}